\documentclass[aps,prd,twocolumn,showpacs,nofootinbib]{revtex4-1}
\usepackage{graphicx}
\usepackage{amsmath}
\usepackage{amssymb}
\usepackage{bm}
\usepackage{bbm}
\usepackage{color}
\usepackage{slashed}
\usepackage[dvipdfm]{hyperref}
\usepackage{hyperref}
\usepackage{stmaryrd}
\usepackage{float}
\begin{document}

\title{Production of the triply heavy $\Omega_{ccc}$ and $\Omega_{bbb}$ baryons at $e^+e^-$ colliders}

\author{Su-Zhi Wu$^{a}$}
\email{zzswu@nwafu.edu.cn}
\author{Pei Wu$^{b}$}
\email{Wupei@huse.edu.cn}
\author{You-Wei Li$^{a}$}
\email{singlebag@nwafu.edu.cn}

\affiliation{ $^a$College of Science, Northwest A$\&$F University, Yangling, Shaanxi 712100, China \\
$^b$College of Science, Hunan University of Science and Engineering, Yongzhou, Hunan 425199, China
}

\begin{abstract}

Nonrelativistic quantum chromodynamics (NRQCD) factorization formulism is an important approach to investigate the production of the heavy quarkonium. In this paper, we study the production of the $\Omega_{ccc}$ and $\Omega_{bbb}$ at the $e^+e^-$ collider, using the NRQCD factorization formulism. We do the full calculation of the total and differential cross sections of the processes $e^+e^-\rightarrow \gamma^*/Z^*\rightarrow \Omega_{QQQ}+\bar{Q}+\bar{Q}+\bar{Q}$, in the leading order at the $e^+e^-$ colliders with different energies. In this paper, $\Omega_{QQQ}$ means the triply heavy baryon $\Omega_{ccc}$ or $\Omega_{bbb}$. And $Q$ denotes the $c$ or $b$ quark. The results show that it is hard to observe them at the $e^+e^-$ colliders directly.
\end{abstract}

\maketitle

\section{Introduction}
The doubly charmed baryon $\Xi_{cc}$ has been observed in both SELEX and LHC \cite{SELEX:2002wqn,LHCb:2017iph}. Then, different decay channels of the $\Xi_{cc}$ were observed in LHC \cite{LHCb:2018pcs,LHCb:2021eaf,LHCb:2022rpd}. The lifetime of the $\Xi_{cc}$ was measured in LHC also \cite{LHCb:2018zpl}. While, none of the triply heavy baryons has been observed in experiment. As the baryonic analogues of the heavy quarkonia, triply heavy baryons are helpful to understand the strong interaction between quarks, the hadron structures and weak decays of the heavy baryons. The triply heavy baryons have been studying in theory. About forty years ago, the heavy baryon spectroscopy was studied using the QCD bag model in \cite{Hasenfratz:1980ka}, firstly. Then, the mass of the triply heavy baryons have been determined in kinds of theoretical models, such as, the QCD sum rules \cite{Zhang:2009re,Wang:2011ae,Wang:2020avt,Aliev:2012tt,Aliev:2014lxa,Alomayrah:2020qyw,Azizi:2014jxa}, the lattice QCD \cite{Meinel:2010pw,Briceno:2012wt,PACS-CS:2013vie,Padmanath:2013zfa,Brown:2014ena,Mathur:2018epb}, the hypercentral quark model \cite{Patel:2008mv, Tazimi:2021ywr,Shah:2019jxp,Rai:2017hue}, the nonrelativistic quark model \cite{Vijande:2015faa,Shah:2017jkr,Yang:2019lsg,Llanes-Estrada:2013rwa,Llanes-Estrada:2011gwu,Kakadiya:2022pin} and the relativistic quark model \cite{Faustov:2021qqf,Martynenko:2007je,Migura:2006ep,Bhavsar:2018tad}, the Faddeev equation \cite{Radin:2014yna,Qin:2019hgk,Zheng:2010zzc}, the symmetry-preserving Schwinger-Dyson equations \cite{Yin:2019bxe,Gutierrez-Guerrero:2019uwa}, the variational method \cite{Jia:2006gw}, the Regge trajectories \cite{Wei:2015gsa,Wei:2016jyk,Oudichhya:2021kop,Oudichhya:2021yln}, etc.

As that two pairs of heavy quarks need to be produced in the production of the doubly heavy baryon \cite{Jin:2013bra,Jiang:2013ej,Jiang:2012jt,Zhang:2011hi,Ma:2003zk,Chen:2014frw,Yang:2014ita}, three pairs of heavy quarks need to be produced in the production progress of the triply heavy baryon. This makes it difficult to
produce and observe the triply heavy baryon in experiment. The production cross sections of the triply heavy baryons at the LHC have been evaluated via fragmentation of the heavy quark in Refs.\cite{Saleev:1999ti,GomshiNobary:2005ur,GomshiNobary:2006tzy,GomshiNobary:2004mq}. In \cite{Chen:2011mb,Wu:2012wj}, the total and differential cross sections of the direct production processes of these baryons at the LHC have been calculated. And the results show that it is quite promising to discover the triply heavy baryons in LHC. The production of the $\Omega_{ccc}$ in heavy-ion collisions have been investigated in Refs.\cite{Zhao:2017gpq,He:2014tga,Becattini:2005hb}. Compared with the hadron collider, the backgrounds at the $e^+e^-$ collider are much less. So, it is necessary to study the production of the triply heavy baryons at the $e^+e^-$ collider. The production of the triply charmed baryon in $e^+e^-$ collisions has been estimated in Ref.\cite{Baranov:2004er}. To simply the calculation, the authors in Ref.\cite{Baranov:2004er} have taken an approximation ignoring the mass of the charm quark in the numerator of the quark propagator and in all traces. However, as pointed out in Ref.\cite{Chang:1992bb}, the approximation will lead to quite large errors. In this work, we report the study on the production of the triply heavy baryons $\Omega_{ccc}$ and $\Omega_{bbb}$ at $e^+e^-$ colliders. We calculate the total and differential cross sections of the direct production of the $\Omega_{ccc}$ and $\Omega_{bbb}$ exactly.

Because the constituent quarks of the triply heavy baryon are all heavy flavored, the NRQCD \cite{Bodwin:1994jh} can be used to describe the triply heavy baryon. The production of the $\Omega_{QQQ}$ can be factorized into two parts, the short-distance coefficient which describes the process of the production of the
three heavy quark pairs, and the long-distance matrix element which
describes the hadronization of three heavy quarks to the triply heavy baryon.
The number of the Feynman diagrams corresponding to the short distance coefficient is large. To achieve the calculation, we utilize the characteristic of the constituent quarks as the authors have done in Refs.\cite{Chen:2011mb,Wu:2012wj}.

The present paper is organized as follows. In Sec.\ref{sec:Production}, we will show the details on the calculation of the two parts, the short-distance coefficient and the long-distance matrix element mentioned above. In Sec.\ref{sec:NANDC}, we will list the numerical results and conclusions. And in Sec.\ref{sumra}, we will give a brief summary.

\section{Production of THE $\Omega_{ccc}$ and $\Omega_{bbb}$ at $e^+e^-$ colliders}\label{sec:Production}
For the triply heavy baryons, the relative motions among the heavy valence quarks are nonrelativistic. The typical velocity $v$ of the heavy quark in the rest frame of the $\Omega_{QQQ}$ is small. And there are three distinct energy scales in the triply heavy baryons, i.e., the mass of the heavy quark $m$, the three-momentum of the heavy quark $mv$, and the energy of the heavy quark $mv^2$. And the three energy scales have the relationship of  $m>>mv>>mv^2$. The direct production of the triply heavy baryons can be described as follows. Three heavy quark pairs are produced firstly at energy scale $m$ or higher followed by the formation of the triply heavy baryon at the energy scale $mv$. Namely, the production cross section of the triply heavy baryon can be factorized into the short-distance coefficient and the long-distance matrix element. The short-distance coefficient describes the process of the three heavy quark pairs production which can be expanded as a power series of $\alpha_s$ and $\alpha$ at the energy scale $m$ or higher. The long-distance matrix element describes the hadronization of the triply heavy baryon from the pointlike three heavy quarks.

 The triply heavy baryon must be in color singlet. In the leading order, the color wave function of the $\Omega_{QQQ}$ must be $\frac{1}{\sqrt{6}}\varepsilon^{\xi_1\xi_2\xi_3}Q_{1\xi_1}Q_{2\xi_2}Q_{3\xi_3}$
 with $\xi_i$ ($i=$1,2,3) being the color index of the valence quark $Q_i$. As a ground state, the orbital angular momentum wave function is symmetrical. The exchange antisymmetry of the identical fermions implies that the $\Omega_{QQQ}$ must be the spin-symmetrical state and
the spin of it must be $\frac{3}{2}$.
 As given in Ref.\cite{Chen:2011mb}, in the leading Fock state description, the \emph{S}-wave state of the nonrelativistic triply heavy baryon $\Omega_{QQQ}$ can be written as,
\begin{eqnarray} \label{nrqcd}
|\Omega_{QQQ},\frac{3}{2},S_Z\rangle&&=\sqrt{2M}\int\frac{d^3\vec{q}_1}{(2\pi)^3}\frac{d^3\vec{q}_2}{(2\pi)^3}\frac{1}{\sqrt{3!}} \nonumber\\
&&\sum_{\xi_1,\xi_2,\xi_3}\sum_{\eta_1,\eta_2,\eta_3}\frac{\varepsilon^{\xi_1\xi_2\xi_3}}{\sqrt{6}}
\langle
\frac{3}{2},S_Z|\eta_1,\eta_2,\eta_3\rangle \nonumber\\
&&\frac{1}{\sqrt{2E_12E_22E_3}}\psi(\vec{q}_1,\vec{q}_2)
|Q_1,\xi_1,\eta_1,\vec{q}_1\rangle\nonumber\\
&&|Q_2,\xi_2,\eta_2,\vec{q}_2\rangle
|Q_3,\xi_3,\eta_3,\vec{q}_3\rangle, \ \ \ \
\end{eqnarray}
\text{where, $\vec{q}_3=-\vec{q}_1-\vec{q}_2$, and} \ \
\begin{eqnarray}
\langle Q_i,\xi_{i},\eta_{i},&&\vec{q}_i|Q_j,\xi_j,\eta_j,\vec{q}_j\rangle \nonumber \\ &&=\delta_{\eta_i\eta_j}\delta_{\xi_i\xi_j}(2\pi)^32E_f\delta^{(3)}(\vec{q}_i-\vec{q}_j),\nonumber
\end{eqnarray}
 with $\eta_i$ and ($E_i$, $\vec{q}_i$)
($i=1,2,3$) being the spin and the four-momentum
of the heavy quark $Q_i$; $M$ being the mass of the baryon $\Omega_{QQQ}$; $\langle \frac{3}{2},S_Z|\eta_1,\eta_2,\eta_3\rangle$
being the Clebsch-Gordan (C-G) coefficient; $S_Z$ being the
third component of the spin of the $\Omega_{ccc}$, and
$\psi(\vec{q}_1,\vec{q}_2)$
being the wave function in the momentum space which is normalized as follows
\begin{eqnarray} \label{nor-wa}
\int\frac{d^3\vec{q}_1}{(2\pi)^3}\frac{d^3\vec{q}_2}{(2\pi)^3}\psi^*(\vec{q}_1,\vec{q}_2)\psi(\vec{q}_1,\vec{q}_2)=1\;.
\end{eqnarray}

In the heavy quark limit, the dependence of the short distance on the momenta, $q_1$ and $q_2$, can be neglected in the leading order. Namely, the momenta of the produced three identical heavy quarks can be treated as the same. As a result, the long-distance matrix is proportional to the wave function of the baryon at the origin,
 \begin{eqnarray} \label{wave}
\Psi(0,0)=\int\frac{d^3q_1}{(2\pi)^3}\frac{d^3q_2}{(2\pi)^3}\psi(q_1,q_2)\;.
\end{eqnarray}

In this paper, we consider only the contribution from the leading order short-distance coefficient, and the leading order long-distance matrix element. By the standard perturbation theory, the amplitude of the process $e^+e^-\rightarrow \Omega_{QQQ}+\bar{Q}+\bar{Q}+\bar{Q}$ can be written as,
 \begin{eqnarray} \label{cross}
&&A(e^+e^-\to\Omega_{QQQ}+\bar{Q}+\bar{Q}+\bar{Q}) = \frac{\sqrt{2M}}{\sqrt{(2m)^3}}
\frac{\Psi(0,0)}{\sqrt{3!}}   \nonumber \\
&&\times \mathcal{M}(e^+e^-\to
(QQQ)_1^{(\frac{3}{2},S_Z)}+\bar{Q}+\bar{Q}+\bar{Q})\;, \ \ \ \
\end{eqnarray}
in which, $\mathcal{M}(e^+e^-\to
(QQQ)_1^{(\frac{3}{2},S_Z)}+\bar{Q}+\bar{Q}+\bar{Q})$ is the matrix element in the leading order of the process $e^+e^-\to
(QQQ)_1^{(\frac{3}{2},S_Z)}+\bar{Q}+\bar{Q}+\bar{Q}$.\footnote{ $(QQQ)_1^{(S,S_Z)}$ means the total spin and the
third component of the  pointlike three $Q$-quarks are $S$ and $S_Z$, respectively; the momenta of the three quarks are the same; and the three heavy quarks couple to a color singlet.} It is the product of the short-distance coefficient with the color- and spin-wave functions in the long-distance matrix element.

\begin{figure*}
\centering
\includegraphics[width=0.42\textwidth]{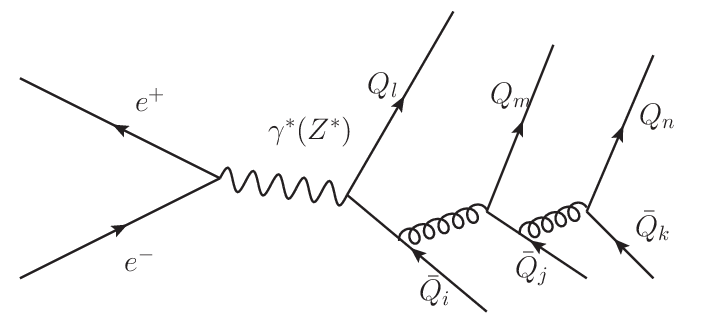}
\includegraphics[width=0.42\textwidth]{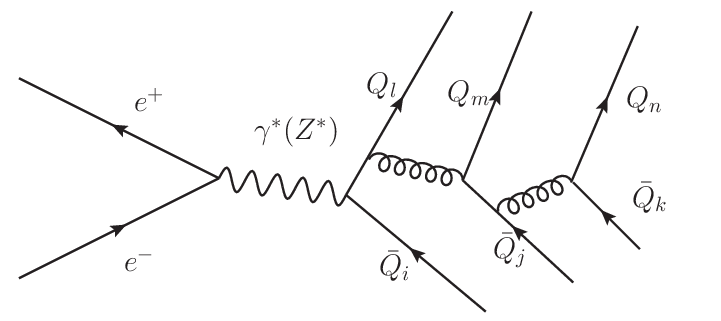}\\
\includegraphics[width=0.42\textwidth]{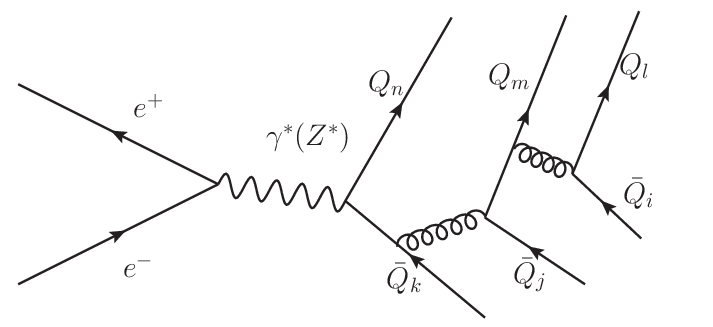}
\includegraphics[width=0.42\textwidth]{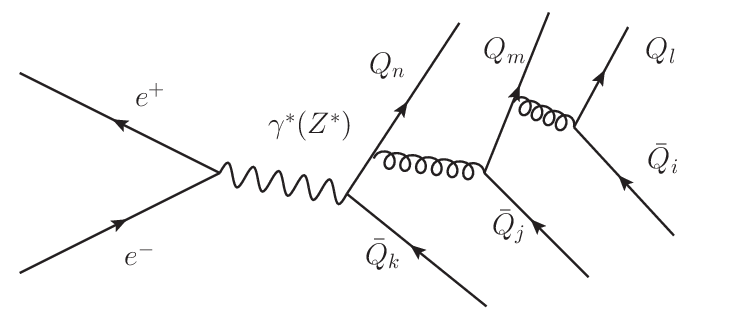}\\
\includegraphics[width=0.42\textwidth]{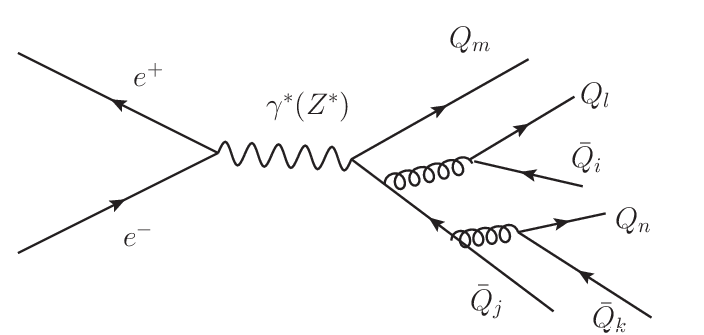}
\includegraphics[width=0.42\textwidth]{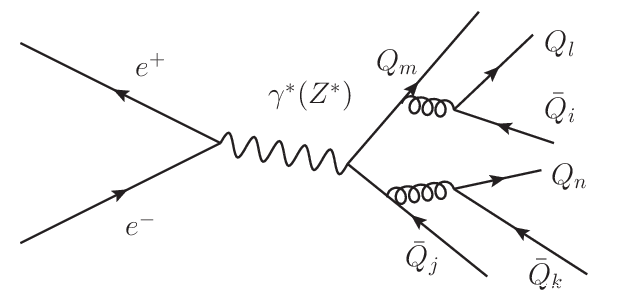}\\
\includegraphics[width=0.42\textwidth]{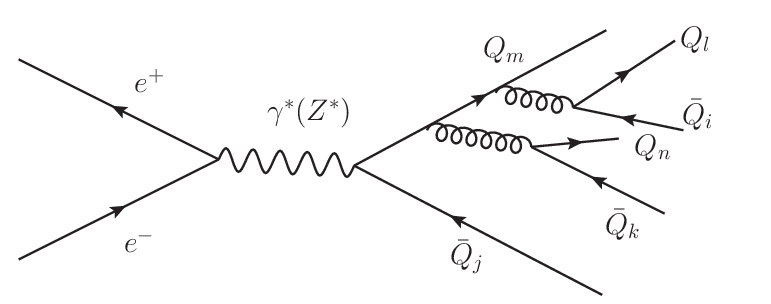}
\includegraphics[width=0.42\textwidth]{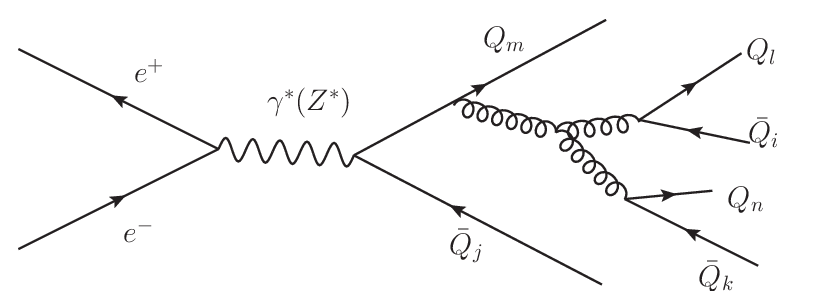}\\
\includegraphics[width=0.42\textwidth]{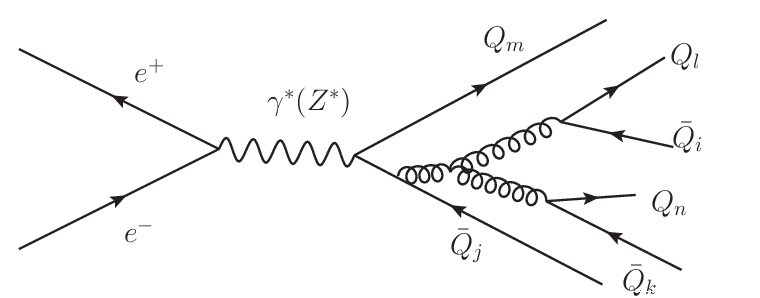}
\caption{Nine typical Feynman diagrams for the process (\ref{anni}). The indices $i,j,k,l,m,n=1,2,3$ (with $i\neq j\neq k$ and $l\neq m\neq n$). }\label{9dth}
\end{figure*}

Now, let us consider the short-distance coefficient. The contribution to the production of the three heavy quark pairs
 comes from the $e^+e^-$ annihilation process mainly,
 \begin{eqnarray}\label{anni}
&&e^-(k_1)+e^+(k_2)\rightarrow Z^*/\gamma^*\rightarrow
Q(p_1,\xi_1)+Q(p_2,\xi_2)\nonumber \\
&&+Q(p_3,\xi_3)+\bar{Q}(p_4,\chi_1)+\bar{Q}(p_5,\chi_2)+\bar{Q}(p_6,\chi_3), \ \ \ \ \ \
\end{eqnarray}
where, $k_1$ and $k_2$ are the 4-momenta of the electron and position;
$p_i$ ($i$=1,6) are the 4-momenta of the produced $Q$ and $\bar{Q}$ quarks, with $p_1=p_2=p_3$; $\xi_j$ and $\chi_j$($j=$1,2,3) are the color indices of the
$Q_{j}$-quarks and $\bar{Q}_{j}$-quarks, respectively. The produced three pairs of heavy quarks in process (\ref{anni}) have six permutations denoted as, $(Q_1\bar{Q}_iQ_2\bar{Q}_jQ_3\bar{Q}_k)$ with $i,j,k$=1,2,3 and $i\neq j\neq k$.\footnote{$(Q_1\bar{Q}_iQ_2\bar{Q}_jQ_3\bar{Q}_k)$ ($i,j,k$=1,2,3 and $i\neq j\neq k$) means the quark $Q_1$ in the same fermion line with the antiquark $\bar{Q}_i$, the quark $Q_2$ in the same fermion line with the antiquark $\bar{Q}_j$ and the quark $Q_3$ in the same fermion line with the antiquark $\bar{Q}_k$.} In the calculation, we will disregard the contribution of the electroweak interaction between the heavy quarks for the process (\ref{anni}). As a result, there are seven inequivalent topology structures for each of the six permutations as the same as the one in \cite{Chen:2011mb}. So, there are 42 topology structures totally for the produced heavy quarks. Inserting the $e^+e^-\gamma^*$ vertex and $e^+e^-Z^*$ vertex into the 42 topology structures in all the allowed positions in the tree level, we get 576 Feynman diagrams for the process (\ref{anni}). And, we show nine typical ones in Fig.\ref{9dth}.

 As pointed out above, the color configuration of the produced three heavy quarks is $\frac{1}{\sqrt{6}}\varepsilon^{\xi_1\xi_2\xi_3}Q_{1\xi_1}Q_{2\xi_2}Q_{3\xi_3}$. Setting $T^a=\frac{\lambda^a}{2}$ with $\lambda^a $ $(a=1,\ldots,8)$ being the Gell-Mann matrices and using the color-flow method in \cite{Chen:2011mb}, we can get the color factors of the 576 Feynman diagrams. The color factors of the last two diagrams in Fig.\ref{9dth} are both,
 \begin{eqnarray}\label{co-fa7}
\sum_{a,b,c}\sum_{\xi_1,\xi_2,\xi_3}\frac{1}{\sqrt{6}}\varepsilon^{\xi_1\xi_2\xi_3}f^{abc}(T^a)_{\xi_l\chi_i}(T^b)_{\xi_m\chi_j}(T^b)_{\xi_n\chi_k}=0\;.\nonumber
\end{eqnarray}
in which, $f^{abc}$ ($(a,b,c=1,\ldots,8)$) is the structure constant of the $SU(3)$ group, and the index $\xi_i$ ($i=1,2,3$) appears twice, in $\varepsilon^{\xi_1\xi_2\xi_3}$ directly and as one of the indices $\xi_l$, $\xi_m$ and $\xi_n$ indirectly. And we see the result is independent onto the indices $i,j,k,l,m,$ and $n$. As a result, we get the conclusion that the total contribution to the amplitude of the process (\ref{anni}) from the 72 Feynman diagrams involving the three-gluon vertex vanishes, because the color factors of these Feynman diagrams are all zero. The number of the Feynman diagrams which we need to consider reduces to 504. The color factors of the first seven Feynman diagrams in Fig.\ref{9dth} are all the same,
\begin{eqnarray}\label{co-fa1}
\sum_{a,b}\sum_{\xi_1,\xi_2,\xi_3}\frac{1}{\sqrt{6}}\varepsilon^{\xi_1\xi_2\xi_3}(T^a)_{\xi_l\chi_i}&&(T^aT^b)_{\xi_m\chi_j}(T^b)_{\xi_n\chi_k} \nonumber \\
&&=(-1)^N\frac{4}{9}\frac{1}{\sqrt{6}}\varepsilon^{\chi_i\chi_j\chi_k}\;, \nonumber
\end{eqnarray}
in which $N$ is the number of permutations of transforming the set $(\xi_l,\xi_m,\xi_n)$ to the set $(\xi_1,\xi_2,\xi_3)$. For each of the Feynman diagrams, there is a Feynman factor $(-1)^{N^*}$, where $N^*$ equals the total number of permutations of transforming the set $(l,m,n)$ to the set $(1,2,3)$ and transforming the set $(i,j,k)$ to the set $(1,2,3)$. Subsuming the Feynman factor into the corresponding color factor, we find that all the color factors of the remaining 504 Feynman diagrams are the same,
\begin{eqnarray}\label{co-fa}
C_{col}=(-1)^{N^*}(-1)^N\frac{4}{9}\frac{1}{\sqrt{6}}\varepsilon^{\chi_i\chi_j\chi_k}=\frac{4}{9}\frac{1}{\sqrt{6}}\varepsilon^{\chi_1\chi_2\chi_3}\;.\nonumber
\end{eqnarray}

\begin{figure}[h]
\centering
\includegraphics[width=0.45\textwidth]{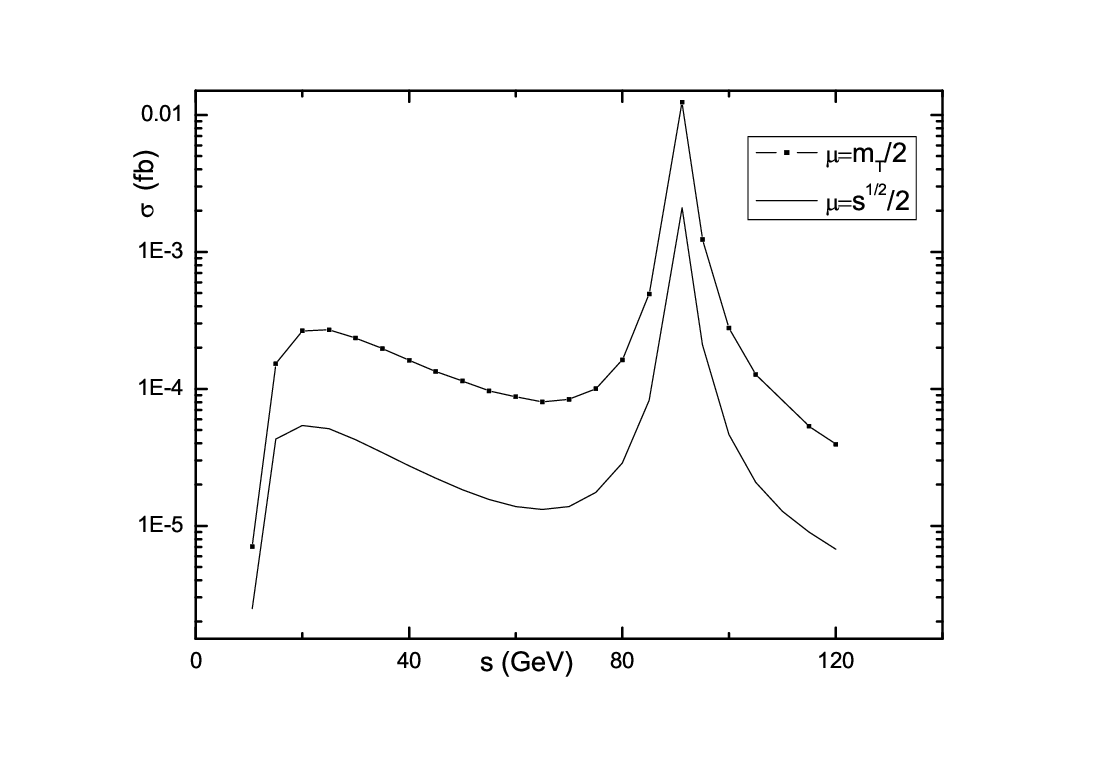}
\caption{The production cross section of the $\Omega_{ccc}$ produced by $e^+e^-$ annihilation at different energies. }\label{3c}
\end{figure}

\begin{figure}[h]
\centering
\includegraphics[width=0.45\textwidth]{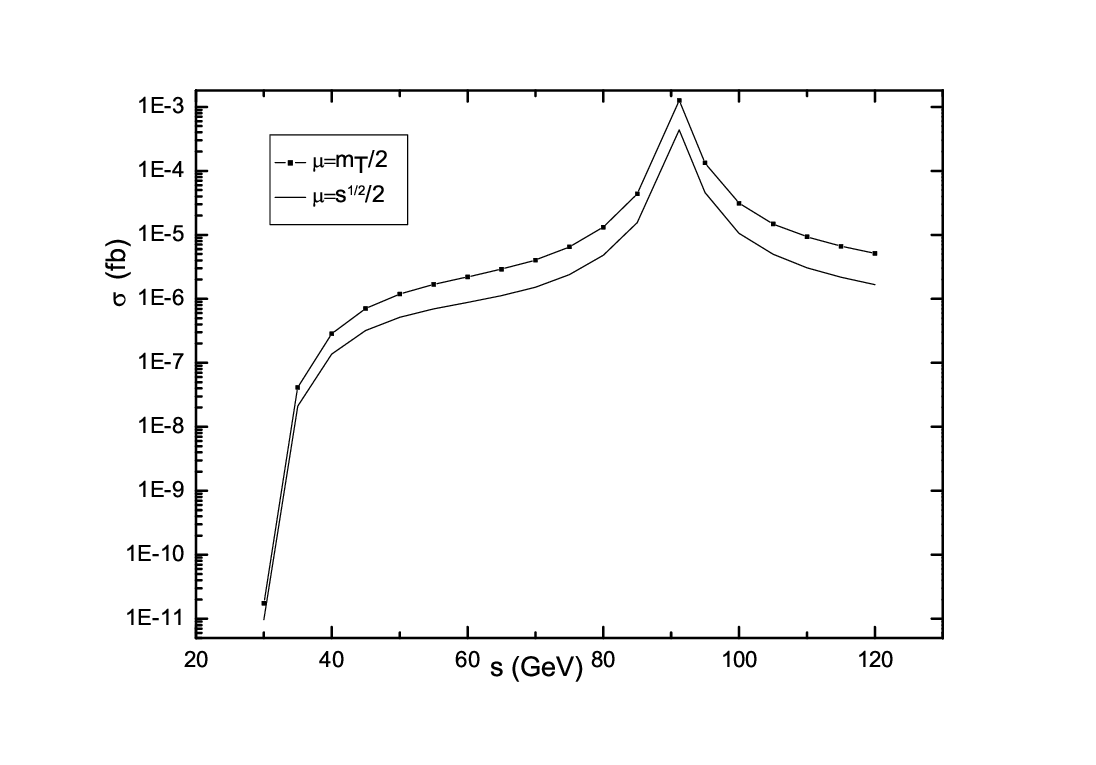}
\caption{The production cross section of the $\Omega_{bbb}$ produced by $e^+e^-$ annihilation at different energies.}\label{3b}
\end{figure}

\begin{figure}[h]
\centering
\includegraphics[width=0.45\textwidth]{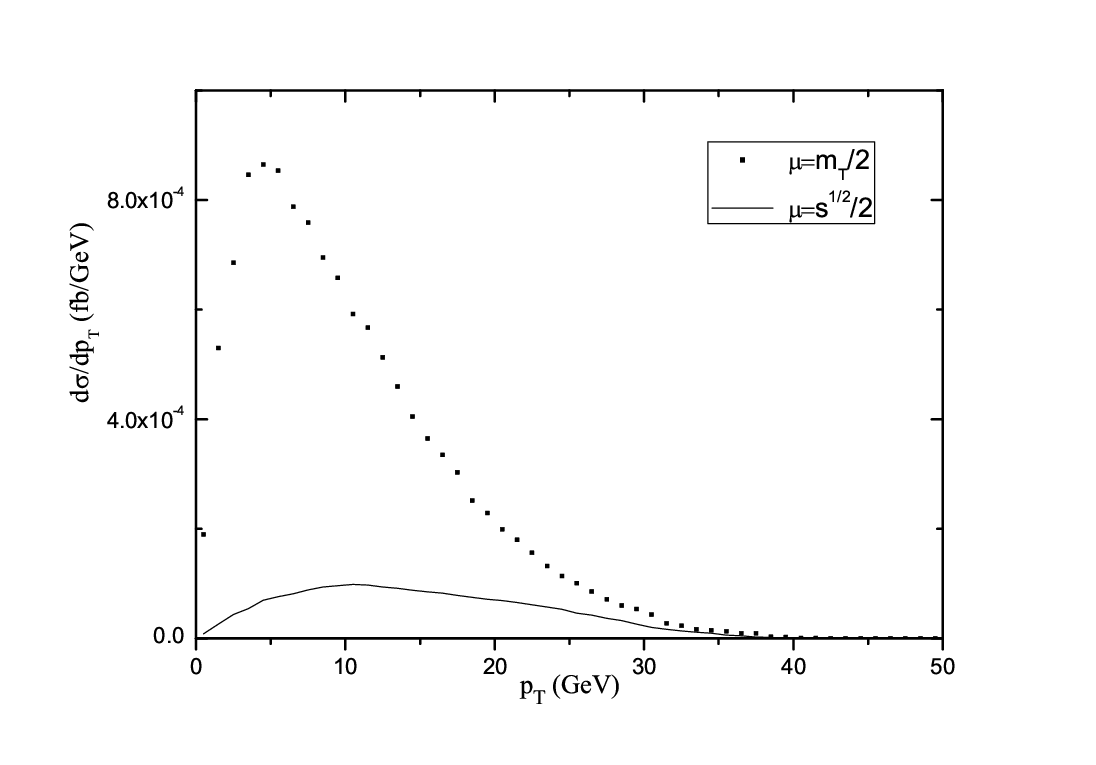}
\caption{The $p_T$ distributions of the $\Omega_{ccc}$ produced at CEPC with $\sqrt{s}=91.2\; {\rm GeV}$. }\label{3cpt}
\end{figure}

\begin{figure}[h]
\centering
\includegraphics[width=0.45\textwidth]{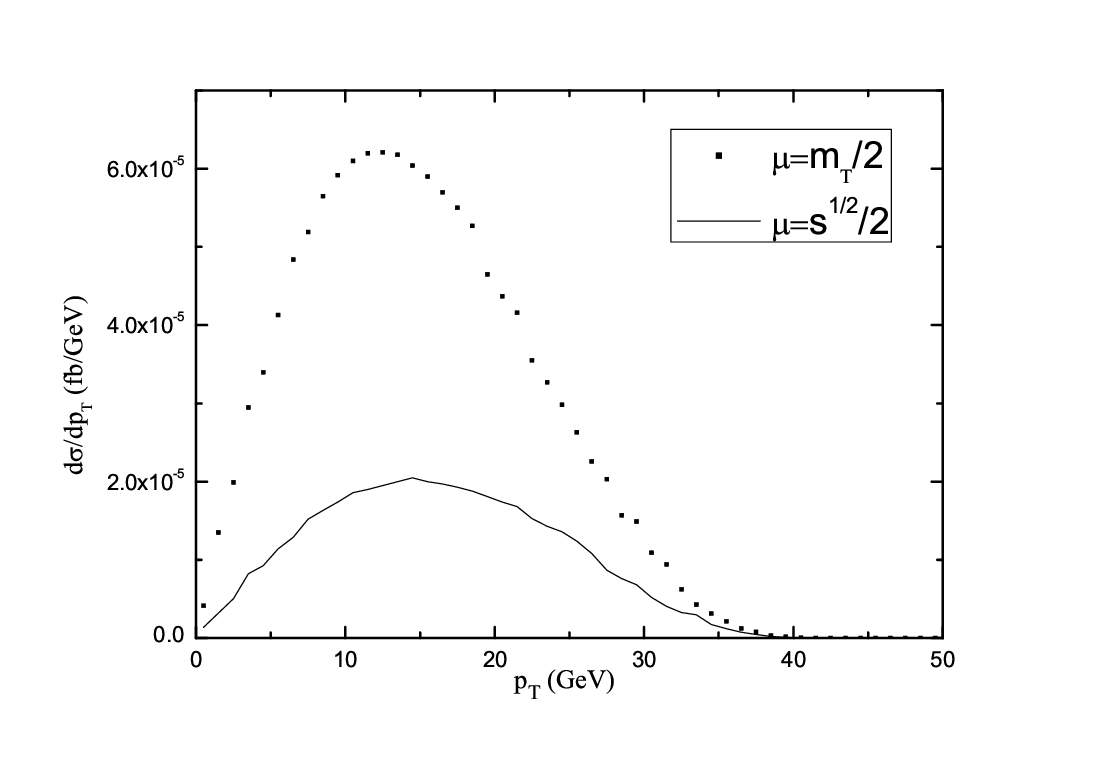}
\caption{The $p_T$ distributions of of the $\Omega_{bbb}$ produced at CEPC with $\sqrt{s}=91.2\; {\rm GeV}$. }\label{3bpt}
\end{figure}

Let us consider the remained 504 Feynman diagrams. For each one of the six permutations given above, there are 84 Feynman diagrams, in which 42 ones correspond to the process $e^-+e^+\to Z^* \to Q_1+Q_2+Q_3+\bar{Q}_1+\bar{Q}_2+\bar{Q}_3$ and the other 42 ones correspond to the process $e^-+e^+\to \gamma^* \to Q_1+Q_2+Q_3+\bar{Q}_1+\bar{Q}_2+\bar{Q}_3$. In the leading order, the momenta of the three heavy quarks are considered to be equal. As a result, the contribution of every 84 Feynam diagrams corresponding to each of the six permutations are the same.  The total amplitude for the process $e^+e^-\to
(QQQ)_1^{(\frac{3}{2},S_Z)}+\bar{Q}+\bar{Q}+\bar{Q}$
\begin{eqnarray} \label{m-c-ee}
  \mathcal{M}(e^+e^-\to
(QQQ)_1^{(\frac{3}{2},S_Z)}&+&\bar{Q}+\bar{Q}+\bar{Q})=\sum_{k=1}^{504}C_{col} \,\Gamma_{k}  \nonumber \\
         &=&3!\sum_{k=1}^{84}C_{col} \,\Gamma^{'}_{k} \ \ ,
\end{eqnarray}
where $\Gamma_{k}$ ($k=1,2....,504$) are the matrices without color factors of the Feynman diagrams, and $\Gamma^{'}_{k}$ ($k=1,2....,84$) are the matrices without color factors of the 84 Feynman diagrams corresponding to the permutation $(Q_1\bar{Q}_1Q_2\bar{Q}_2Q_3\bar{Q}_3)$.

\section{Numerical results and discussions}\label{sec:NANDC}

Now, the cross section of the progress (\ref{anni}) can be written as
\begin{eqnarray}
 \sigma=&&\frac{1}{3!}\int\sum_{S_Z,\varsigma_i}\frac{(2\pi)^4}{2\hat{s}}
  \delta^4(k_1+k_2-P-p_4-p_5-p_6)
\nonumber\\
 &&d\Pi_4\frac{1}{4}\sum_{s_{1},s_{2},\chi_i}|\mathcal{A}(e^+e^-\to
  \Omega_{QQQ}\bar{Q}\bar{Q}\bar{Q})|^2,
\end{eqnarray}
with
\begin{eqnarray}
d\Pi_4=\frac{d^3P}{(2\pi)^32E}\frac{d^3p_4}{(2\pi)^32E_{p_4}}
   \frac{d^3p_5}{(2\pi)^32E_{p_5}}
   \frac{d^3p_6}{(2\pi)^32E_{p_6}},
\end{eqnarray}
where $\varsigma_i$ ($i=1,2,3$), $s_1$ and $s_2$ are the spins of the antiquarks $\bar{Q}_i$, $e^+$ and $e^-$, respectively.

To do the numerical calculation, the parameters are taken as follows: $m_Z=91.18\,{\rm GeV}$,  $\Gamma_{_Z}=2.49\,{\rm GeV}$, $\alpha(m_Z)=1/127.95$, $sin^2\theta_w=0.224$, $m_c=1.5\,{\rm GeV}$, $|\Psi_{\Omega_{ccc}}(0,0)|^2=0.36\cdot10^{-3}\,{\rm GeV^6}$ as in \cite{Baranov:2004er}, and $m_b=4.9\,{\rm GeV}$, $|\Psi_{\Omega_{bbb}}(0,0)|^2=0.189\,{\rm GeV^6}$ as in \cite{Wu:2012wj}.

For the electromagnetic coupling constant, we adopt
\begin{eqnarray}
  \alpha(q)&=&\frac{\alpha(m_Z)}{1-\frac{2\alpha(m_Z)}{3\pi}{\rm log}(\frac{q}{m_Z})},
\end{eqnarray}
 where, $q$ denotes the energy scale of the electromagnetic coupling constant, and we take $q=\frac{\sqrt{s}}{2}$ with $\sqrt{s}$ being the colliding energy in the center-of-mass frame. And the strong coupling constant, we adopt
\begin{eqnarray}
  \alpha_s(\mu)&=&\frac{\alpha_s(m_Z)}{1+\frac{b_0}{2\pi}\alpha_s(m_Z){\rm log}(\frac{\mu}{m_Z})},\\
  \text{with} \ \    b_0&=&11-\frac{2}{3}n_f,\nonumber
\end{eqnarray}
where $\mu$ is the energy scale and
$\alpha_s(m_Z)=0.118$. For the production of the $\Omega_{ccc}$, $n_f$ is taken to be 4 when $\mu\leq 2m_b$
and 5 when $\mu$ is larger than 2$m_b$. And for the production of the $\Omega_{bbb}$, $n_f$ is taken to be 5. For comparison, we take two
different values for $\mu$, i.e., $\mu$=$\sqrt{s}/2$ and $\mu$=$m_T/2$, where $m_T$ is the transverse mass of the produced baryon, namely
$m_T^2$=$p_T^2$+$M^2$.\footnote{Here, $p_T$ means the transverse momentum of the produced triply heavy baryon.} And in our numerical calculation, the integrals of the phase-pace are achieved by VEGAS \cite{Lepage:1977sw}.

SuperKEKB, the asymmetric beam energy $e^-e^+$ collider, is an upgrade of the KEKB accelerator facility. The target integrated luminosity of 50 $ab^{-1}$ to be collected by the Belle II experiment. And a high luminosity $e^+e^-$ collider named the Circular Electron-Positron Collider (CEPC) has been proposed by the Chinese particle physics community. The CEPC is designed to operate at three different modes, as a Higgs factory at $\sqrt{s}=240\,{\rm GeV}$, as a $Z$ factory at $\sqrt{s}=91.2\,{\rm GeV}$ and performing $WW$ threshold scans around $\sqrt{s}=160\,{\rm GeV}$ \cite{An:2018dwb}. We calculate the production cross sections of the baryons $\Omega_{ccc}$ and $\Omega_{bbb}$ at CEPC with the energy $\sqrt{s}=91.2\,{\rm GeV}$ and $\sqrt{s}=160\,{\rm GeV}$, and at Belle II with the center-of-mass energy $\sqrt{s}=10.58\,{\rm GeV}$. The results are shown in Table \ref{cross1t}. And, we calculate the production of the $\Omega_{ccc}$ and $\Omega_{bbb}$ at the $e^+e^-$ colliders with different colliding energies, which are shown in Fig.\ref{3c} and Fig.\ref{3b}. From the Table \ref{cross1t}, Fig.\ref{3c} and Fig.\ref{3b}, we see that the cross sections for the production of the $\Omega_{ccc}$ and $\Omega_{bbb}$ at $\sqrt{s}=91.2\,{\rm GeV}$ are larger than the corresponding ones at other colliding energies. This is because the cross sections contributed by the $Z$-boson exchange processes are proportional to a factor of $1/[(s-m_{_Z}^2)^2+m_{_Z}^2 \Gamma_Z^2]$, which has a peak at the $Z$ pole and decreases rapidly when the $\sqrt{s}$ deviates from $m_Z$, as pointed in Ref.\cite{Zhang:2021ypo}.

\begin{table}[htb]
\begin{tabular}{|c|c|c|c|c|}
\hline
-&$\sqrt{s}$&$\mu$&$\Omega_{ccc}$&$\Omega_{bbb}$ \\
\hline
CEPC&91.2 {\rm GeV}& $\sqrt{s}/2$&0.00204(5)&$0.437(6)\times10^{-3}$\\
-&-&$m_T/2$&0.0124(3)&$1.25(2)\times10^{-3}$\\
-&160 {\rm GeV}& $\sqrt{s}/2$&$0.214(9)\times10^{-5}$&$0.55(2)\times10^{-6}$\\
-&-&$m_T/2$&$0.108(9)\times10^{-4}$&$0.183(4)\times10^{-5}$\\
\hline
Belle II&10.58 {\rm GeV}& $\sqrt{s}/2$&$0.249(1)\times10^{-5}$&-\\
-&-&$m_T/2$&$0.707(3)\times10^{-5}$&-\\
\hline
\end{tabular}
\caption{Production cross sections (in unit $fb$) of the $\Omega_{ccc}$ and $\Omega_{bbb}$ at the CEPC and Belle II. }
\label{cross1t}
\end{table}

In this paper, we also calculate the differential cross sections $d \sigma/dp_T$ of the production of the $\Omega_{ccc}$ and $\Omega_{bbb}$ at $\sqrt{s}=91.2 \ \ {\rm GeV}$, which are shown in Figs.\ref{3cpt} and \ref{3bpt}.

As we see that both the integrated cross sections and differential cross
 sections are proportional to $|\Psi(0,0)|^2$, $\alpha^2(q)$and $\alpha_s^4(\mu)$.
 So the numerical results can be changed by one even two orders using
different the wave function at the origin of the triply heavy
baryons, different running coupling constants and different energy
scale choices. We can get the conclusion from the numerical results shown in Figs.\ref{3c} and \ref{3b}, that the production cross sections of the $\Omega_{ccc}$ and $\Omega_{bbb}$ take their maximum values at $\sqrt{s}=m_Z$.
From the numerical results in Table \ref{cross1t}, we know that it is impossible to observe $\Omega_{ccc}$ at SuperKEKB, if there is no other models to produce them at the $e^+e^-$ colliders. Also, we see there are around $32-198$ events for the production of $\Omega_{ccc}$ and $7-20$ events for the production of $\Omega_{bbb}$, at CEPC with $16 \,ab^{-1}$ integrated luminosity running at $91.2 \, {\rm GeV}$.

In this work, we also calculate the production cross section of the $\Omega_{ccc}$ using the same parameters as in Ref. \cite{Baranov:2004er}. And the numerical result is $0.20(2)\times10^{-2}\, {\rm fb}$, which is about an order of magnitude smaller than the one $0.0404(4)\, {\rm fb}$ in Ref.\cite{Baranov:2004er}. This huge difference may can be explained by the approximation adopted in Ref.\cite{Baranov:2004er}.

\section{Summary}\label{sumra}
We have studied the production of the $\Omega_{ccc}$ and $\Omega_{bbb}$ at the SuperKEKB and the CEPC. From the numerical results, we conclude that it is impossible to find the triply heavy baryons at SuperKEKB, and it is hard to find $\Omega_{ccc}$ and $\Omega_{bbb}$ at the CEPC because of the few events.

\hspace{1cm}

\noindent {\bf Acknowledgments:} The authors thank Doctor Jian-Wei Xu for helpful discussions and important suggestions on the manuscript. The
work of S. Z. Wu was supported by the National Nature Science Foundation of China under Grants No. 11347200.


\begin{thebibliography}{66}%
\makeatletter
\providecommand \@ifxundefined [1]{%
 \@ifx{#1\undefined}
}%
\providecommand \@ifnum [1]{%
 \ifnum #1\expandafter \@firstoftwo
 \else \expandafter \@secondoftwo
 \fi
}%
\providecommand \@ifx [1]{%
 \ifx #1\expandafter \@firstoftwo
 \else \expandafter \@secondoftwo
 \fi
}%
\providecommand \natexlab [1]{#1}%
\providecommand \enquote  [1]{``#1''}%
\providecommand \bibnamefont  [1]{#1}%
\providecommand \bibfnamefont [1]{#1}%
\providecommand \citenamefont [1]{#1}%
\providecommand \href@noop [0]{\@secondoftwo}%
\providecommand \href [0]{\begingroup \@sanitize@url \@href}%
\providecommand \@href[1]{\@@startlink{#1}\@@href}%
\providecommand \@@href[1]{\endgroup#1\@@endlink}%
\providecommand \@sanitize@url [0]{\catcode `\\12\catcode `\$12\catcode
  `\&12\catcode `\#12\catcode `\^12\catcode `\_12\catcode `\%12\relax}%
\providecommand \@@startlink[1]{}%
\providecommand \@@endlink[0]{}%
\providecommand \url  [0]{\begingroup\@sanitize@url \@url }%
\providecommand \@url [1]{\endgroup\@href {#1}{\urlprefix }}%
\providecommand \urlprefix  [0]{URL }%
\providecommand \Eprint [0]{\href }%
\providecommand \doibase [0]{http://dx.doi.org/}%
\providecommand \selectlanguage [0]{\@gobble}%
\providecommand \bibinfo  [0]{\@secondoftwo}%
\providecommand \bibfield  [0]{\@secondoftwo}%
\providecommand \translation [1]{[#1]}%
\providecommand \BibitemOpen [0]{}%
\providecommand \bibitemStop [0]{}%
\providecommand \bibitemNoStop [0]{.\EOS\space}%
\providecommand \EOS [0]{\spacefactor3000\relax}%
\providecommand \BibitemShut  [1]{\csname bibitem#1\endcsname}%
\let\auto@bib@innerbib\@empty
\bibitem [{\citenamefont {Mattson}\ \emph {et~al.}(2002)\citenamefont {Mattson}
  \emph {et~al.}}]{SELEX:2002wqn}%
  \BibitemOpen
  \bibfield  {author} {\bibinfo {author} {\bibfnamefont {M.}~\bibnamefont
  {Mattson}} \emph {et~al.} (\bibinfo {collaboration} {SELEX}),\ }\bibfield
  {title} {\emph {\bibinfo {title} {{First Observation of the Doubly Charmed
  Baryon $\Xi^+_{cc}$}},\ }}\href {\doibase 10.1103/PhysRevLett.89.112001}
  {\bibfield  {journal} {\bibinfo  {journal} {Phys. Rev. Lett.}\ }\textbf
  {\bibinfo {volume} {89}},\ \bibinfo {pages} {112001} (\bibinfo {year}
  {2002})},\ \Eprint {http://arxiv.org/abs/hep-ex/0208014}
  {arXiv:hep-ex/0208014} \BibitemShut {NoStop}%
\bibitem [{\citenamefont {Aaij}\ \emph {et~al.}(2017)\citenamefont {Aaij} \emph
  {et~al.}}]{LHCb:2017iph}%
  \BibitemOpen
  \bibfield  {author} {\bibinfo {author} {\bibfnamefont {R.}~\bibnamefont
  {Aaij}} \emph {et~al.} (\bibinfo {collaboration} {LHCb}),\ }\bibfield
  {title} {\emph {\bibinfo {title} {{Observation of the doubly charmed baryon
  $\Xi_{cc}^{++}$}},\ }}\href {\doibase 10.1103/PhysRevLett.119.112001}
  {\bibfield  {journal} {\bibinfo  {journal} {Phys. Rev. Lett.}\ }\textbf
  {\bibinfo {volume} {119}},\ \bibinfo {pages} {112001} (\bibinfo {year}
  {2017})},\ \Eprint {http://arxiv.org/abs/1707.01621} {arXiv:1707.01621
  [hep-ex]} \BibitemShut {NoStop}%
\bibitem [{\citenamefont {Aaij}\ \emph
  {et~al.}(2018{\natexlab{a}})\citenamefont {Aaij} \emph
  {et~al.}}]{LHCb:2018pcs}%
  \BibitemOpen
  \bibfield  {author} {\bibinfo {author} {\bibfnamefont {R.}~\bibnamefont
  {Aaij}} \emph {et~al.} (\bibinfo {collaboration} {LHCb}),\ }\bibfield
  {title} {\emph {\bibinfo {title} {{First Observation of the Doubly Charmed
  Baryon Decay $\Xi_{cc}^{++}\rightarrow \Xi_{c}^{+}\pi^{+}$}},\ }}\href
  {\doibase 10.1103/PhysRevLett.121.162002} {\bibfield  {journal} {\bibinfo
  {journal} {Phys. Rev. Lett.}\ }\textbf {\bibinfo {volume} {121}},\ \bibinfo
  {pages} {162002} (\bibinfo {year} {2018}{\natexlab{a}})},\ \Eprint
  {http://arxiv.org/abs/1807.01919} {arXiv:1807.01919 [hep-ex]} \BibitemShut
  {NoStop}%
\bibitem [{\citenamefont {Aaij}\ \emph {et~al.}(2021)\citenamefont {Aaij} \emph
  {et~al.}}]{LHCb:2021eaf}%
  \BibitemOpen
  \bibfield  {author} {\bibinfo {author} {\bibfnamefont {R.}~\bibnamefont
  {Aaij}} \emph {et~al.} (\bibinfo {collaboration} {LHCb}),\ }\bibfield
  {title} {\emph {\bibinfo {title} {{Search for the doubly charmed baryon $
  {\varXi}_{cc}^{+} $ in the $ {\varXi}_c^{+}{\pi}^{-}{\pi}^{+} $ final
  state}},\ }}\href {\doibase 10.1007/JHEP12(2021)107} {\bibfield  {journal}
  {\bibinfo  {journal} {JHEP}\ }\textbf {\bibinfo {volume} {12}},\ \bibinfo
  {pages} {107} (\bibinfo {year} {2021})},\ \Eprint
  {http://arxiv.org/abs/2109.07292} {arXiv:2109.07292 [hep-ex]} \BibitemShut
  {NoStop}%
\bibitem [{\citenamefont {Aaij}\ \emph {et~al.}(2022)\citenamefont {Aaij} \emph
  {et~al.}}]{LHCb:2022rpd}%
  \BibitemOpen
  \bibfield  {author} {\bibinfo {author} {\bibfnamefont {R.}~\bibnamefont
  {Aaij}} \emph {et~al.} (\bibinfo {collaboration} {LHCb}),\ }\bibfield
  {title} {\emph {\bibinfo {title} {{Observation of the doubly charmed baryon
  decay $ {\varXi}_{cc}^{++}\to {\varXi}_c^{\prime +}{\pi}^{+} $}},\ }}\href
  {\doibase 10.1007/JHEP05(2022)038} {\bibfield  {journal} {\bibinfo  {journal}
  {JHEP}\ }\textbf {\bibinfo {volume} {05}},\ \bibinfo {pages} {038} (\bibinfo
  {year} {2022})},\ \Eprint {http://arxiv.org/abs/2202.05648} {arXiv:2202.05648
  [hep-ex]} \BibitemShut {NoStop}%
\bibitem [{\citenamefont {Aaij}\ \emph
  {et~al.}(2018{\natexlab{b}})\citenamefont {Aaij} \emph
  {et~al.}}]{LHCb:2018zpl}%
  \BibitemOpen
  \bibfield  {author} {\bibinfo {author} {\bibfnamefont {R.}~\bibnamefont
  {Aaij}} \emph {et~al.} (\bibinfo {collaboration} {LHCb}),\ }\bibfield
  {title} {\emph {\bibinfo {title} {{Measurement of the Lifetime of the Doubly
  Charmed Baryon $\Xi_{cc}^{++}$}},\ }}\href {\doibase
  10.1103/PhysRevLett.121.052002} {\bibfield  {journal} {\bibinfo  {journal}
  {Phys. Rev. Lett.}\ }\textbf {\bibinfo {volume} {121}},\ \bibinfo {pages}
  {052002} (\bibinfo {year} {2018}{\natexlab{b}})},\ \Eprint
  {http://arxiv.org/abs/1806.02744} {arXiv:1806.02744 [hep-ex]} \BibitemShut
  {NoStop}%
\bibitem [{\citenamefont {Hasenfratz}\ \emph {et~al.}(1980)\citenamefont
  {Hasenfratz}, \citenamefont {Horgan}, \citenamefont {Kuti},\ and\
  \citenamefont {Richard}}]{Hasenfratz:1980ka}%
  \BibitemOpen
  \bibfield  {author} {\bibinfo {author} {\bibfnamefont {P.}~\bibnamefont
  {Hasenfratz}}, \bibinfo {author} {\bibfnamefont {R.~R.}\ \bibnamefont
  {Horgan}}, \bibinfo {author} {\bibfnamefont {J.}~\bibnamefont {Kuti}}, \ and\
  \bibinfo {author} {\bibfnamefont {J.~M.}\ \bibnamefont {Richard}},\
  }\bibfield  {title} {\emph {\bibinfo {title} {{Heavy Baryon Spectroscopy in
  the {QCD} Bag Model}},\ }}\href {\doibase 10.1016/0370-2693(80)90906-5}
  {\bibfield  {journal} {\bibinfo  {journal} {Phys. Lett. B}\ }\textbf
  {\bibinfo {volume} {94}},\ \bibinfo {pages} {401} (\bibinfo {year}
  {1980})}\BibitemShut {NoStop}%
\bibitem [{\citenamefont {Zhang}\ and\ \citenamefont
  {Huang}(2009)}]{Zhang:2009re}%
  \BibitemOpen
  \bibfield  {author} {\bibinfo {author} {\bibfnamefont {J.-R.}\ \bibnamefont
  {Zhang}}\ and\ \bibinfo {author} {\bibfnamefont {M.-Q.}\ \bibnamefont
  {Huang}},\ }\bibfield  {title} {\emph {\bibinfo {title} {{Deciphering triply
  heavy baryons in terms of QCD sum rules}},\ }}\href {\doibase
  10.1016/j.physletb.2009.02.056} {\bibfield  {journal} {\bibinfo  {journal}
  {Phys. Lett. B}\ }\textbf {\bibinfo {volume} {674}},\ \bibinfo {pages} {28}
  (\bibinfo {year} {2009})},\ \Eprint {http://arxiv.org/abs/0902.3297}
  {arXiv:0902.3297 [hep-ph]} \BibitemShut {NoStop}%
\bibitem [{\citenamefont {Wang}(2012)}]{Wang:2011ae}%
  \BibitemOpen
  \bibfield  {author} {\bibinfo {author} {\bibfnamefont {Z.-G.}\ \bibnamefont
  {Wang}},\ }\bibfield  {title} {\emph {\bibinfo {title} {{Analysis of the
  Triply Heavy Baryon States with QCD Sum Rules}},\ }}\href {\doibase
  10.1088/0253-6102/58/5/17} {\bibfield  {journal} {\bibinfo  {journal}
  {Commun. Theor. Phys.}\ }\textbf {\bibinfo {volume} {58}},\ \bibinfo {pages}
  {723} (\bibinfo {year} {2012})},\ \Eprint {http://arxiv.org/abs/1112.2274}
  {arXiv:1112.2274 [hep-ph]} \BibitemShut {NoStop}%
\bibitem [{\citenamefont {Wang}(2021)}]{Wang:2020avt}%
  \BibitemOpen
  \bibfield  {author} {\bibinfo {author} {\bibfnamefont {Z.-G.}\ \bibnamefont
  {Wang}},\ }\bibfield  {title} {\emph {\bibinfo {title} {{Analysis of the
  triply-heavy baryon states with the QCD sum rules}},\ }}\href {\doibase
  10.1007/s43673-021-00006-3} {\bibfield  {journal} {\bibinfo  {journal} {AAPPS
  Bull.}\ }\textbf {\bibinfo {volume} {31}},\ \bibinfo {pages} {5} (\bibinfo
  {year} {2021})},\ \Eprint {http://arxiv.org/abs/2010.08939} {arXiv:2010.08939
  [hep-ph]} \BibitemShut {NoStop}%
\bibitem [{\citenamefont {Aliev}\ \emph {et~al.}(2013)\citenamefont {Aliev},
  \citenamefont {Azizi},\ and\ \citenamefont {Savci}}]{Aliev:2012tt}%
  \BibitemOpen
  \bibfield  {author} {\bibinfo {author} {\bibfnamefont {T.~M.}\ \bibnamefont
  {Aliev}}, \bibinfo {author} {\bibfnamefont {K.}~\bibnamefont {Azizi}}, \ and\
  \bibinfo {author} {\bibfnamefont {M.}~\bibnamefont {Savci}},\ }\bibfield
  {title} {\emph {\bibinfo {title} {{Masses and Residues of the Triply Heavy
  Spin-1/2 Baryons}},\ }}\href {\doibase 10.1007/JHEP04(2013)042} {\bibfield
  {journal} {\bibinfo  {journal} {JHEP}\ }\textbf {\bibinfo {volume} {04}},\
  \bibinfo {pages} {042} (\bibinfo {year} {2013})},\ \Eprint
  {http://arxiv.org/abs/1212.6065} {arXiv:1212.6065 [hep-ph]} \BibitemShut
  {NoStop}%
\bibitem [{\citenamefont {Aliev}\ \emph {et~al.}(2014)\citenamefont {Aliev},
  \citenamefont {Azizi},\ and\ \citenamefont {Savc\i{}}}]{Aliev:2014lxa}%
  \BibitemOpen
  \bibfield  {author} {\bibinfo {author} {\bibfnamefont {T.~M.}\ \bibnamefont
  {Aliev}}, \bibinfo {author} {\bibfnamefont {K.}~\bibnamefont {Azizi}}, \ and\
  \bibinfo {author} {\bibfnamefont {M.}~\bibnamefont {Savc\i{}}},\ }\bibfield
  {title} {\emph {\bibinfo {title} {{Properties of triply heavy spin-3/2
  baryons}},\ }}\href {\doibase 10.1088/0954-3899/41/6/065003} {\bibfield
  {journal} {\bibinfo  {journal} {J. Phys. G}\ }\textbf {\bibinfo {volume}
  {41}},\ \bibinfo {pages} {065003} (\bibinfo {year} {2014})},\ \Eprint
  {http://arxiv.org/abs/1404.2091} {arXiv:1404.2091 [hep-ph]} \BibitemShut
  {NoStop}%
\bibitem [{\citenamefont {Alomayrah}\ and\ \citenamefont
  {Barakat}(2020)}]{Alomayrah:2020qyw}%
  \BibitemOpen
  \bibfield  {author} {\bibinfo {author} {\bibfnamefont {N.}~\bibnamefont
  {Alomayrah}}\ and\ \bibinfo {author} {\bibfnamefont {T.}~\bibnamefont
  {Barakat}},\ }\bibfield  {title} {\emph {\bibinfo {title} {{The excited
  states of triply-heavy baryons in QCD sum rules}},\ }}\href {\doibase
  10.1140/epja/s10050-020-00062-7} {\bibfield  {journal} {\bibinfo  {journal}
  {Eur. Phys. J. A}\ }\textbf {\bibinfo {volume} {56}},\ \bibinfo {pages} {76}
  (\bibinfo {year} {2020})}\BibitemShut {NoStop}%
\bibitem [{\citenamefont {Azizi}\ \emph {et~al.}(2014)\citenamefont {Azizi},
  \citenamefont {Aliev},\ and\ \citenamefont {Savci}}]{Azizi:2014jxa}%
  \BibitemOpen
  \bibfield  {author} {\bibinfo {author} {\bibfnamefont {K.}~\bibnamefont
  {Azizi}}, \bibinfo {author} {\bibfnamefont {T.~M.}\ \bibnamefont {Aliev}}, \
  and\ \bibinfo {author} {\bibfnamefont {M.}~\bibnamefont {Savci}},\ }\bibfield
   {title} {\emph {\bibinfo {title} {{Properties of doubly and triply heavy
  baryons}},\ }}\href {\doibase 10.1088/1742-6596/556/1/012016} {\bibfield
  {journal} {\bibinfo  {journal} {J. Phys. Conf. Ser.}\ }\textbf {\bibinfo
  {volume} {556}},\ \bibinfo {pages} {012016} (\bibinfo {year}
  {2014})}\BibitemShut {NoStop}%
\bibitem [{\citenamefont {Meinel}(2010)}]{Meinel:2010pw}%
  \BibitemOpen
  \bibfield  {author} {\bibinfo {author} {\bibfnamefont {S.}~\bibnamefont
  {Meinel}},\ }\bibfield  {title} {\emph {\bibinfo {title} {{Prediction of the
  $Omega_{bbb}$ mass from lattice QCD}},\ }}\href {\doibase
  10.1103/PhysRevD.82.114514} {\bibfield  {journal} {\bibinfo  {journal} {Phys.
  Rev. D}\ }\textbf {\bibinfo {volume} {82}},\ \bibinfo {pages} {114514}
  (\bibinfo {year} {2010})},\ \Eprint {http://arxiv.org/abs/1008.3154}
  {arXiv:1008.3154 [hep-lat]} \BibitemShut {NoStop}%
\bibitem [{\citenamefont {Briceno}\ \emph {et~al.}(2012)\citenamefont
  {Briceno}, \citenamefont {Lin},\ and\ \citenamefont
  {Bolton}}]{Briceno:2012wt}%
  \BibitemOpen
  \bibfield  {author} {\bibinfo {author} {\bibfnamefont {R.~A.}\ \bibnamefont
  {Briceno}}, \bibinfo {author} {\bibfnamefont {H.-W.}\ \bibnamefont {Lin}}, \
  and\ \bibinfo {author} {\bibfnamefont {D.~R.}\ \bibnamefont {Bolton}},\
  }\bibfield  {title} {\emph {\bibinfo {title} {{Charmed-Baryon Spectroscopy
  from Lattice QCD with $N_f$ = 2+1+1 Flavors}},\ }}\href {\doibase
  10.1103/PhysRevD.86.094504} {\bibfield  {journal} {\bibinfo  {journal} {Phys.
  Rev. D}\ }\textbf {\bibinfo {volume} {86}},\ \bibinfo {pages} {094504}
  (\bibinfo {year} {2012})},\ \Eprint {http://arxiv.org/abs/1207.3536}
  {arXiv:1207.3536 [hep-lat]} \BibitemShut {NoStop}%
\bibitem [{\citenamefont {Namekawa}\ \emph {et~al.}(2013)\citenamefont
  {Namekawa} \emph {et~al.}}]{PACS-CS:2013vie}%
  \BibitemOpen
  \bibfield  {author} {\bibinfo {author} {\bibfnamefont {Y.}~\bibnamefont
  {Namekawa}} \emph {et~al.} (\bibinfo {collaboration} {PACS-CS}),\ }\bibfield
  {title} {\emph {\bibinfo {title} {{Charmed baryons at the physical point in
  2+1 flavor lattice QCD}},\ }}\href {\doibase 10.1103/PhysRevD.87.094512}
  {\bibfield  {journal} {\bibinfo  {journal} {Phys. Rev. D}\ }\textbf {\bibinfo
  {volume} {87}},\ \bibinfo {pages} {094512} (\bibinfo {year} {2013})},\
  \Eprint {http://arxiv.org/abs/1301.4743} {arXiv:1301.4743 [hep-lat]}
  \BibitemShut {NoStop}%
\bibitem [{\citenamefont {Padmanath}\ \emph {et~al.}(2014)\citenamefont
  {Padmanath}, \citenamefont {Edwards}, \citenamefont {Mathur},\ and\
  \citenamefont {Peardon}}]{Padmanath:2013zfa}%
  \BibitemOpen
  \bibfield  {author} {\bibinfo {author} {\bibfnamefont {M.}~\bibnamefont
  {Padmanath}}, \bibinfo {author} {\bibfnamefont {R.~G.}\ \bibnamefont
  {Edwards}}, \bibinfo {author} {\bibfnamefont {N.}~\bibnamefont {Mathur}}, \
  and\ \bibinfo {author} {\bibfnamefont {M.}~\bibnamefont {Peardon}},\
  }\bibfield  {title} {\emph {\bibinfo {title} {{Spectroscopy of triply-charmed
  baryons from lattice QCD}},\ }}\href {\doibase 10.1103/PhysRevD.90.074504}
  {\bibfield  {journal} {\bibinfo  {journal} {Phys. Rev. D}\ }\textbf {\bibinfo
  {volume} {90}},\ \bibinfo {pages} {074504} (\bibinfo {year} {2014})},\
  \Eprint {http://arxiv.org/abs/1307.7022} {arXiv:1307.7022 [hep-lat]}
  \BibitemShut {NoStop}%
\bibitem [{\citenamefont {Brown}\ \emph {et~al.}(2014)\citenamefont {Brown},
  \citenamefont {Detmold}, \citenamefont {Meinel},\ and\ \citenamefont
  {Orginos}}]{Brown:2014ena}%
  \BibitemOpen
  \bibfield  {author} {\bibinfo {author} {\bibfnamefont {Z.~S.}\ \bibnamefont
  {Brown}}, \bibinfo {author} {\bibfnamefont {W.}~\bibnamefont {Detmold}},
  \bibinfo {author} {\bibfnamefont {S.}~\bibnamefont {Meinel}}, \ and\ \bibinfo
  {author} {\bibfnamefont {K.}~\bibnamefont {Orginos}},\ }\bibfield  {title}
  {\emph {\bibinfo {title} {{Charmed bottom baryon spectroscopy from lattice
  QCD}},\ }}\href {\doibase 10.1103/PhysRevD.90.094507} {\bibfield  {journal}
  {\bibinfo  {journal} {Phys. Rev. D}\ }\textbf {\bibinfo {volume} {90}},\
  \bibinfo {pages} {094507} (\bibinfo {year} {2014})},\ \Eprint
  {http://arxiv.org/abs/1409.0497} {arXiv:1409.0497 [hep-lat]} \BibitemShut
  {NoStop}%
\bibitem [{\citenamefont {Mathur}\ \emph {et~al.}(2018)\citenamefont {Mathur},
  \citenamefont {Padmanath},\ and\ \citenamefont {Mondal}}]{Mathur:2018epb}%
  \BibitemOpen
  \bibfield  {author} {\bibinfo {author} {\bibfnamefont {N.}~\bibnamefont
  {Mathur}}, \bibinfo {author} {\bibfnamefont {M.}~\bibnamefont {Padmanath}}, \
  and\ \bibinfo {author} {\bibfnamefont {S.}~\bibnamefont {Mondal}},\
  }\bibfield  {title} {\emph {\bibinfo {title} {{Precise predictions of
  charmed-bottom hadrons from lattice QCD}},\ }}\href {\doibase
  10.1103/PhysRevLett.121.202002} {\bibfield  {journal} {\bibinfo  {journal}
  {Phys. Rev. Lett.}\ }\textbf {\bibinfo {volume} {121}},\ \bibinfo {pages}
  {202002} (\bibinfo {year} {2018})},\ \Eprint
  {http://arxiv.org/abs/1806.04151} {arXiv:1806.04151 [hep-lat]} \BibitemShut
  {NoStop}%
\bibitem [{\citenamefont {Patel}\ \emph {et~al.}(2009)\citenamefont {Patel},
  \citenamefont {Majethiya},\ and\ \citenamefont {Vinodkumar}}]{Patel:2008mv}%
  \BibitemOpen
  \bibfield  {author} {\bibinfo {author} {\bibfnamefont {B.}~\bibnamefont
  {Patel}}, \bibinfo {author} {\bibfnamefont {A.}~\bibnamefont {Majethiya}}, \
  and\ \bibinfo {author} {\bibfnamefont {P.~C.}\ \bibnamefont {Vinodkumar}},\
  }\bibfield  {title} {\emph {\bibinfo {title} {{Masses and Magnetic moments of
  Triply Heavy Flavour Baryons in Hypercentral Model}},\ }}\href {\doibase
  10.1007/s12043-009-0061-4} {\bibfield  {journal} {\bibinfo  {journal}
  {Pramana}\ }\textbf {\bibinfo {volume} {72}},\ \bibinfo {pages} {679}
  (\bibinfo {year} {2009})},\ \Eprint {http://arxiv.org/abs/0808.2880}
  {arXiv:0808.2880 [hep-ph]} \BibitemShut {NoStop}%
\bibitem [{\citenamefont {Tazimi}\ and\ \citenamefont
  {Ghasempour}(2021)}]{Tazimi:2021ywr}%
  \BibitemOpen
  \bibfield  {author} {\bibinfo {author} {\bibfnamefont {N.}~\bibnamefont
  {Tazimi}}\ and\ \bibinfo {author} {\bibfnamefont {A.}~\bibnamefont
  {Ghasempour}},\ }\bibfield  {title} {\emph {\bibinfo {title} {{Mass spectrum
  of triply heavy baryon in the hypercentral quark model}},\ }}\href {\doibase
  10.1142/S0217732321502709} {\bibfield  {journal} {\bibinfo  {journal} {Mod.
  Phys. Lett. A}\ }\textbf {\bibinfo {volume} {36}},\ \bibinfo {pages}
  {2150270} (\bibinfo {year} {2021})}\BibitemShut {NoStop}%
\bibitem [{\citenamefont {Shah}\ and\ \citenamefont
  {Rai}(2019)}]{Shah:2019jxp}%
  \BibitemOpen
  \bibfield  {author} {\bibinfo {author} {\bibfnamefont {Z.}~\bibnamefont
  {Shah}}\ and\ \bibinfo {author} {\bibfnamefont {A.~K.}\ \bibnamefont {Rai}},\
  }\bibfield  {title} {\emph {\bibinfo {title} {{Mass spectra of triply heavy
  charm-beauty baryons}},\ }}\href {\doibase 10.1051/epjconf/201920206001}
  {\bibfield  {journal} {\bibinfo  {journal} {EPJ Web Conf.}\ }\textbf
  {\bibinfo {volume} {202}},\ \bibinfo {pages} {06001} (\bibinfo {year}
  {2019})}\BibitemShut {NoStop}%
\bibitem [{\citenamefont {Rai}\ and\ \citenamefont {Shah}(2017)}]{Rai:2017hue}%
  \BibitemOpen
  \bibfield  {author} {\bibinfo {author} {\bibfnamefont {A.~K.}\ \bibnamefont
  {Rai}}\ and\ \bibinfo {author} {\bibfnamefont {Z.}~\bibnamefont {Shah}},\
  }\bibfield  {title} {\emph {\bibinfo {title} {{Regge Trajectories of triply
  heavy baryons}},\ }}\href {\doibase 10.1088/1742-6596/934/1/012035}
  {\bibfield  {journal} {\bibinfo  {journal} {J. Phys. Conf. Ser.}\ }\textbf
  {\bibinfo {volume} {934}},\ \bibinfo {pages} {012035} (\bibinfo {year}
  {2017})}\BibitemShut {NoStop}%
\bibitem [{\citenamefont {Vijande}\ \emph {et~al.}(2015)\citenamefont
  {Vijande}, \citenamefont {Valcarce},\ and\ \citenamefont
  {Garcilazo}}]{Vijande:2015faa}%
  \BibitemOpen
  \bibfield  {author} {\bibinfo {author} {\bibfnamefont {J.}~\bibnamefont
  {Vijande}}, \bibinfo {author} {\bibfnamefont {A.}~\bibnamefont {Valcarce}}, \
  and\ \bibinfo {author} {\bibfnamefont {H.}~\bibnamefont {Garcilazo}},\
  }\bibfield  {title} {\emph {\bibinfo {title} {{Constituent-quark model
  description of triply heavy baryon nonperturbative lattice QCD data}},\
  }}\href {\doibase 10.1103/PhysRevD.91.054011} {\bibfield  {journal} {\bibinfo
   {journal} {Phys. Rev. D}\ }\textbf {\bibinfo {volume} {91}},\ \bibinfo
  {pages} {054011} (\bibinfo {year} {2015})},\ \Eprint
  {http://arxiv.org/abs/1507.03735} {arXiv:1507.03735 [hep-ph]} \BibitemShut
  {NoStop}%
\bibitem [{\citenamefont {Shah}\ and\ \citenamefont
  {Rai}(2017)}]{Shah:2017jkr}%
  \BibitemOpen
  \bibfield  {author} {\bibinfo {author} {\bibfnamefont {Z.}~\bibnamefont
  {Shah}}\ and\ \bibinfo {author} {\bibfnamefont {A.~K.}\ \bibnamefont {Rai}},\
  }\bibfield  {title} {\emph {\bibinfo {title} {{Masses and Regge trajectories
  of triply heavy $\Omega_{ccc}$ and $\Omega_{bbb}$ baryons}},\ }}\href
  {\doibase 10.1140/epja/i2017-12386-2} {\bibfield  {journal} {\bibinfo
  {journal} {Eur. Phys. J. A}\ }\textbf {\bibinfo {volume} {53}},\ \bibinfo
  {pages} {195} (\bibinfo {year} {2017})}\BibitemShut {NoStop}%
\bibitem [{\citenamefont {Yang}\ \emph {et~al.}(2020)\citenamefont {Yang},
  \citenamefont {Ping}, \citenamefont {Ortega},\ and\ \citenamefont
  {Segovia}}]{Yang:2019lsg}%
  \BibitemOpen
  \bibfield  {author} {\bibinfo {author} {\bibfnamefont {G.}~\bibnamefont
  {Yang}}, \bibinfo {author} {\bibfnamefont {J.}~\bibnamefont {Ping}}, \bibinfo
  {author} {\bibfnamefont {P.~G.}\ \bibnamefont {Ortega}}, \ and\ \bibinfo
  {author} {\bibfnamefont {J.}~\bibnamefont {Segovia}},\ }\bibfield  {title}
  {\emph {\bibinfo {title} {{Triply heavy baryons in the constituent quark
  model}},\ }}\href {\doibase 10.1088/1674-1137/44/2/023102} {\bibfield
  {journal} {\bibinfo  {journal} {Chin. Phys. C}\ }\textbf {\bibinfo {volume}
  {44}},\ \bibinfo {pages} {023102} (\bibinfo {year} {2020})},\ \Eprint
  {http://arxiv.org/abs/1904.10166} {arXiv:1904.10166 [hep-ph]} \BibitemShut
  {NoStop}%
\bibitem [{\citenamefont {Llanes-Estrada}\ \emph {et~al.}(2013)\citenamefont
  {Llanes-Estrada}, \citenamefont {Pavlova},\ and\ \citenamefont
  {Williams}}]{Llanes-Estrada:2013rwa}%
  \BibitemOpen
  \bibfield  {author} {\bibinfo {author} {\bibfnamefont {F.~J.}\ \bibnamefont
  {Llanes-Estrada}}, \bibinfo {author} {\bibfnamefont {O.~I.}\ \bibnamefont
  {Pavlova}}, \ and\ \bibinfo {author} {\bibfnamefont {R.}~\bibnamefont
  {Williams}},\ }\bibfield  {title} {\emph {\bibinfo {title} {{Triply heavy
  baryon mass estimated within pNRQCD}},\ }}\href {\doibase
  10.5506/APhysPolBSupp.6.821} {\bibfield  {journal} {\bibinfo  {journal} {Acta
  Phys. Polon. Supp.}\ }\textbf {\bibinfo {volume} {6}},\ \bibinfo {pages}
  {821} (\bibinfo {year} {2013})},\ \Eprint {http://arxiv.org/abs/1304.3636}
  {arXiv:1304.3636 [nucl-th]} \BibitemShut {NoStop}%
\bibitem [{\citenamefont {Llanes-Estrada}\ \emph {et~al.}(2012)\citenamefont
  {Llanes-Estrada}, \citenamefont {Pavlova},\ and\ \citenamefont
  {Williams}}]{Llanes-Estrada:2011gwu}%
  \BibitemOpen
  \bibfield  {author} {\bibinfo {author} {\bibfnamefont {F.~J.}\ \bibnamefont
  {Llanes-Estrada}}, \bibinfo {author} {\bibfnamefont {O.~I.}\ \bibnamefont
  {Pavlova}}, \ and\ \bibinfo {author} {\bibfnamefont {R.}~\bibnamefont
  {Williams}},\ }\bibfield  {title} {\emph {\bibinfo {title} {{A First Estimate
  of Triply Heavy Baryon Masses from the pNRQCD Perturbative Static
  Potential}},\ }}\href {\doibase 10.1140/epjc/s10052-012-2019-9} {\bibfield
  {journal} {\bibinfo  {journal} {Eur. Phys. J. C}\ }\textbf {\bibinfo {volume}
  {72}},\ \bibinfo {pages} {2019} (\bibinfo {year} {2012})},\ \Eprint
  {http://arxiv.org/abs/1111.7087} {arXiv:1111.7087 [hep-ph]} \BibitemShut
  {NoStop}%
\bibitem [{\citenamefont {Kakadiya}\ \emph {et~al.}(2022)\citenamefont
  {Kakadiya}, \citenamefont {Shah},\ and\ \citenamefont
  {Rai}}]{Kakadiya:2022pin}%
  \BibitemOpen
  \bibfield  {author} {\bibinfo {author} {\bibfnamefont {A.}~\bibnamefont
  {Kakadiya}}, \bibinfo {author} {\bibfnamefont {Z.}~\bibnamefont {Shah}}, \
  and\ \bibinfo {author} {\bibfnamefont {A.~K.}\ \bibnamefont {Rai}},\
  }\bibfield  {title} {\emph {\bibinfo {title} {{Spectroscopy of
  \ensuremath{\Omega}ccc and \ensuremath{\Omega}bbb baryons}},\ }}\href
  {\doibase 10.1142/S0217751X22502256} {\bibfield  {journal} {\bibinfo
  {journal} {Int. J. Mod. Phys. A}\ }\textbf {\bibinfo {volume} {37}},\
  \bibinfo {pages} {2250225} (\bibinfo {year} {2022})},\ \Eprint
  {http://arxiv.org/abs/2303.03771} {arXiv:2303.03771 [hep-ph]} \BibitemShut
  {NoStop}%
\bibitem [{\citenamefont {Faustov}\ and\ \citenamefont
  {Galkin}(2022)}]{Faustov:2021qqf}%
  \BibitemOpen
  \bibfield  {author} {\bibinfo {author} {\bibfnamefont {R.~N.}\ \bibnamefont
  {Faustov}}\ and\ \bibinfo {author} {\bibfnamefont {V.~O.}\ \bibnamefont
  {Galkin}},\ }\bibfield  {title} {\emph {\bibinfo {title} {{Triply heavy
  baryon spectroscopy in the relativistic quark model}},\ }}\href {\doibase
  10.1103/PhysRevD.105.014013} {\bibfield  {journal} {\bibinfo  {journal}
  {Phys. Rev. D}\ }\textbf {\bibinfo {volume} {105}},\ \bibinfo {pages}
  {014013} (\bibinfo {year} {2022})},\ \Eprint
  {http://arxiv.org/abs/2111.07702} {arXiv:2111.07702 [hep-ph]} \BibitemShut
  {NoStop}%
\bibitem [{\citenamefont {Martynenko}(2008)}]{Martynenko:2007je}%
  \BibitemOpen
  \bibfield  {author} {\bibinfo {author} {\bibfnamefont {A.~P.}\ \bibnamefont
  {Martynenko}},\ }\bibfield  {title} {\emph {\bibinfo {title} {{Ground-state
  triply and doubly heavy baryons in a relativistic three-quark model}},\
  }}\href {\doibase 10.1016/j.physletb.2008.04.030} {\bibfield  {journal}
  {\bibinfo  {journal} {Phys. Lett. B}\ }\textbf {\bibinfo {volume} {663}},\
  \bibinfo {pages} {317} (\bibinfo {year} {2008})},\ \Eprint
  {http://arxiv.org/abs/0708.2033} {arXiv:0708.2033 [hep-ph]} \BibitemShut
  {NoStop}%
\bibitem [{\citenamefont {Migura}\ \emph {et~al.}(2006)\citenamefont {Migura},
  \citenamefont {Merten}, \citenamefont {Metsch},\ and\ \citenamefont
  {Petry}}]{Migura:2006ep}%
  \BibitemOpen
  \bibfield  {author} {\bibinfo {author} {\bibfnamefont {S.}~\bibnamefont
  {Migura}}, \bibinfo {author} {\bibfnamefont {D.}~\bibnamefont {Merten}},
  \bibinfo {author} {\bibfnamefont {B.}~\bibnamefont {Metsch}}, \ and\ \bibinfo
  {author} {\bibfnamefont {H.-R.}\ \bibnamefont {Petry}},\ }\bibfield  {title}
  {\emph {\bibinfo {title} {{Charmed baryons in a relativistic quark model}},\
  }}\href {\doibase 10.1140/epja/i2006-10017-9} {\bibfield  {journal} {\bibinfo
   {journal} {Eur. Phys. J. A}\ }\textbf {\bibinfo {volume} {28}},\ \bibinfo
  {pages} {41} (\bibinfo {year} {2006})},\ \Eprint
  {http://arxiv.org/abs/hep-ph/0602153} {arXiv:hep-ph/0602153} \BibitemShut
  {NoStop}%
\bibitem [{\citenamefont {Bhavsar}\ \emph {et~al.}(2018)\citenamefont
  {Bhavsar}, \citenamefont {Shah},\ and\ \citenamefont
  {Vinodkumar}}]{Bhavsar:2018tad}%
  \BibitemOpen
  \bibfield  {author} {\bibinfo {author} {\bibfnamefont {T.}~\bibnamefont
  {Bhavsar}}, \bibinfo {author} {\bibfnamefont {M.}~\bibnamefont {Shah}}, \
  and\ \bibinfo {author} {\bibfnamefont {P.~C.}\ \bibnamefont {Vinodkumar}},\
  }\bibfield  {title} {\emph {\bibinfo {title} {{A relativistic approach for
  triply heavy avour baryon}},\ }}\href@noop {} {\bibfield  {journal} {\bibinfo
   {journal} {DAE Symp. Nucl. Phys.}\ }\textbf {\bibinfo {volume} {63}},\
  \bibinfo {pages} {840} (\bibinfo {year} {2018})}\BibitemShut {NoStop}%
\bibitem [{\citenamefont {Radin}\ \emph {et~al.}(2014)\citenamefont {Radin},
  \citenamefont {Babaghodrat},\ and\ \citenamefont
  {Monemzadeh}}]{Radin:2014yna}%
  \BibitemOpen
  \bibfield  {author} {\bibinfo {author} {\bibfnamefont {M.}~\bibnamefont
  {Radin}}, \bibinfo {author} {\bibfnamefont {S.}~\bibnamefont {Babaghodrat}},
  \ and\ \bibinfo {author} {\bibfnamefont {M.}~\bibnamefont {Monemzadeh}},\
  }\bibfield  {title} {\emph {\bibinfo {title} {{Estimation of heavy baryon
  masses \ensuremath{\Omega}ccc++ and \ensuremath{\Omega}bbb- by solving the
  Faddeev equation in a three-dimensional approach}},\ }}\href {\doibase
  10.1103/PhysRevD.90.047701} {\bibfield  {journal} {\bibinfo  {journal} {Phys.
  Rev. D}\ }\textbf {\bibinfo {volume} {90}},\ \bibinfo {pages} {047701}
  (\bibinfo {year} {2014})}\BibitemShut {NoStop}%
\bibitem [{\citenamefont {Qin}\ \emph {et~al.}(2019)\citenamefont {Qin},
  \citenamefont {Roberts},\ and\ \citenamefont {Schmidt}}]{Qin:2019hgk}%
  \BibitemOpen
  \bibfield  {author} {\bibinfo {author} {\bibfnamefont {S.-x.}\ \bibnamefont
  {Qin}}, \bibinfo {author} {\bibfnamefont {C.~D.}\ \bibnamefont {Roberts}}, \
  and\ \bibinfo {author} {\bibfnamefont {S.~M.}\ \bibnamefont {Schmidt}},\
  }\bibfield  {title} {\emph {\bibinfo {title} {{Spectrum of light- and
  heavy-baryons}},\ }}\href {\doibase 10.1007/s00601-019-1488-x} {\bibfield
  {journal} {\bibinfo  {journal} {Few Body Syst.}\ }\textbf {\bibinfo {volume}
  {60}},\ \bibinfo {pages} {26} (\bibinfo {year} {2019})},\ \Eprint
  {http://arxiv.org/abs/1902.00026} {arXiv:1902.00026 [nucl-th]} \BibitemShut
  {NoStop}%
\bibitem [{\citenamefont {Zheng}\ and\ \citenamefont
  {Pang}(2010)}]{Zheng:2010zzc}%
  \BibitemOpen
  \bibfield  {author} {\bibinfo {author} {\bibfnamefont {W.}~\bibnamefont
  {Zheng}}\ and\ \bibinfo {author} {\bibfnamefont {H.~R.}\ \bibnamefont
  {Pang}},\ }\bibfield  {title} {\emph {\bibinfo {title} {{Momentum-space
  Faddeev calculations for ground-state triply and doubly heavy baryons in the
  constituent quark model}},\ }}\href {\doibase 10.1142/S0217732310032962}
  {\bibfield  {journal} {\bibinfo  {journal} {Mod. Phys. Lett. A}\ }\textbf
  {\bibinfo {volume} {25}},\ \bibinfo {pages} {2077} (\bibinfo {year}
  {2010})}\BibitemShut {NoStop}%
\bibitem [{\citenamefont {Yin}\ \emph {et~al.}(2019)\citenamefont {Yin},
  \citenamefont {Chen}, \citenamefont {Krein}, \citenamefont {Roberts},
  \citenamefont {Segovia},\ and\ \citenamefont {Xu}}]{Yin:2019bxe}%
  \BibitemOpen
  \bibfield  {author} {\bibinfo {author} {\bibfnamefont {P.-L.}\ \bibnamefont
  {Yin}}, \bibinfo {author} {\bibfnamefont {C.}~\bibnamefont {Chen}}, \bibinfo
  {author} {\bibfnamefont {G.~a.}\ \bibnamefont {Krein}}, \bibinfo {author}
  {\bibfnamefont {C.~D.}\ \bibnamefont {Roberts}}, \bibinfo {author}
  {\bibfnamefont {J.}~\bibnamefont {Segovia}}, \ and\ \bibinfo {author}
  {\bibfnamefont {S.-S.}\ \bibnamefont {Xu}},\ }\bibfield  {title} {\emph
  {\bibinfo {title} {{Masses of ground-state mesons and baryons, including
  those with heavy quarks}},\ }}\href {\doibase 10.1103/PhysRevD.100.034008}
  {\bibfield  {journal} {\bibinfo  {journal} {Phys. Rev. D}\ }\textbf {\bibinfo
  {volume} {100}},\ \bibinfo {pages} {034008} (\bibinfo {year} {2019})},\
  \Eprint {http://arxiv.org/abs/1903.00160} {arXiv:1903.00160 [nucl-th]}
  \BibitemShut {NoStop}%
\bibitem [{\citenamefont {Guti\'errez-Guerrero}\ \emph
  {et~al.}(2019)\citenamefont {Guti\'errez-Guerrero}, \citenamefont {Bashir},
  \citenamefont {Bedolla},\ and\ \citenamefont
  {Santopinto}}]{Gutierrez-Guerrero:2019uwa}%
  \BibitemOpen
  \bibfield  {author} {\bibinfo {author} {\bibfnamefont {L.~X.}\ \bibnamefont
  {Guti\'errez-Guerrero}}, \bibinfo {author} {\bibfnamefont {A.}~\bibnamefont
  {Bashir}}, \bibinfo {author} {\bibfnamefont {M.~A.}\ \bibnamefont {Bedolla}},
  \ and\ \bibinfo {author} {\bibfnamefont {E.}~\bibnamefont {Santopinto}},\
  }\bibfield  {title} {\emph {\bibinfo {title} {{Masses of Light and Heavy
  Mesons and Baryons: A Unified Picture}},\ }}\href {\doibase
  10.1103/PhysRevD.100.114032} {\bibfield  {journal} {\bibinfo  {journal}
  {Phys. Rev. D}\ }\textbf {\bibinfo {volume} {100}},\ \bibinfo {pages}
  {114032} (\bibinfo {year} {2019})},\ \Eprint
  {http://arxiv.org/abs/1911.09213} {arXiv:1911.09213 [nucl-th]} \BibitemShut
  {NoStop}%
\bibitem [{\citenamefont {Jia}(2006)}]{Jia:2006gw}%
  \BibitemOpen
  \bibfield  {author} {\bibinfo {author} {\bibfnamefont {Y.}~\bibnamefont
  {Jia}},\ }\bibfield  {title} {\emph {\bibinfo {title} {{Variational study of
  weakly coupled triply heavy baryons}},\ }}\href {\doibase
  10.1088/1126-6708/2006/10/073} {\bibfield  {journal} {\bibinfo  {journal}
  {JHEP}\ }\textbf {\bibinfo {volume} {10}},\ \bibinfo {pages} {073} (\bibinfo
  {year} {2006})},\ \Eprint {http://arxiv.org/abs/hep-ph/0607290}
  {arXiv:hep-ph/0607290} \BibitemShut {NoStop}%
\bibitem [{\citenamefont {Wei}\ \emph {et~al.}(2015)\citenamefont {Wei},
  \citenamefont {Chen},\ and\ \citenamefont {Guo}}]{Wei:2015gsa}%
  \BibitemOpen
  \bibfield  {author} {\bibinfo {author} {\bibfnamefont {K.-W.}\ \bibnamefont
  {Wei}}, \bibinfo {author} {\bibfnamefont {B.}~\bibnamefont {Chen}}, \ and\
  \bibinfo {author} {\bibfnamefont {X.-H.}\ \bibnamefont {Guo}},\ }\bibfield
  {title} {\emph {\bibinfo {title} {{Masses of doubly and triply charmed
  baryons}},\ }}\href {\doibase 10.1103/PhysRevD.92.076008} {\bibfield
  {journal} {\bibinfo  {journal} {Phys. Rev. D}\ }\textbf {\bibinfo {volume}
  {92}},\ \bibinfo {pages} {076008} (\bibinfo {year} {2015})},\ \Eprint
  {http://arxiv.org/abs/1503.05184} {arXiv:1503.05184 [hep-ph]} \BibitemShut
  {NoStop}%
\bibitem [{\citenamefont {Wei}\ \emph {et~al.}(2017)\citenamefont {Wei},
  \citenamefont {Chen}, \citenamefont {Liu}, \citenamefont {Wang},\ and\
  \citenamefont {Guo}}]{Wei:2016jyk}%
  \BibitemOpen
  \bibfield  {author} {\bibinfo {author} {\bibfnamefont {K.-W.}\ \bibnamefont
  {Wei}}, \bibinfo {author} {\bibfnamefont {B.}~\bibnamefont {Chen}}, \bibinfo
  {author} {\bibfnamefont {N.}~\bibnamefont {Liu}}, \bibinfo {author}
  {\bibfnamefont {Q.-Q.}\ \bibnamefont {Wang}}, \ and\ \bibinfo {author}
  {\bibfnamefont {X.-H.}\ \bibnamefont {Guo}},\ }\bibfield  {title} {\emph
  {\bibinfo {title} {{Spectroscopy of singly, doubly, and triply bottom
  baryons}},\ }}\href {\doibase 10.1103/PhysRevD.95.116005} {\bibfield
  {journal} {\bibinfo  {journal} {Phys. Rev. D}\ }\textbf {\bibinfo {volume}
  {95}},\ \bibinfo {pages} {116005} (\bibinfo {year} {2017})},\ \Eprint
  {http://arxiv.org/abs/1609.02512} {arXiv:1609.02512 [hep-ph]} \BibitemShut
  {NoStop}%
\bibitem [{\citenamefont {Oudichhya}\ \emph
  {et~al.}(2021{\natexlab{a}})\citenamefont {Oudichhya}, \citenamefont
  {Gandhi},\ and\ \citenamefont {Rai}}]{Oudichhya:2021kop}%
  \BibitemOpen
  \bibfield  {author} {\bibinfo {author} {\bibfnamefont {J.}~\bibnamefont
  {Oudichhya}}, \bibinfo {author} {\bibfnamefont {K.}~\bibnamefont {Gandhi}}, \
  and\ \bibinfo {author} {\bibfnamefont {A.~K.}\ \bibnamefont {Rai}},\
  }\bibfield  {title} {\emph {\bibinfo {title} {{Ground and excited state
  masses of \ensuremath{\Omega}c0, \ensuremath{\Omega}cc+ and
  \ensuremath{\Omega}ccc++ baryons}},\ }}\href {\doibase
  10.1103/PhysRevD.103.114030} {\bibfield  {journal} {\bibinfo  {journal}
  {Phys. Rev. D}\ }\textbf {\bibinfo {volume} {103}},\ \bibinfo {pages}
  {114030} (\bibinfo {year} {2021}{\natexlab{a}})},\ \Eprint
  {http://arxiv.org/abs/2105.10647} {arXiv:2105.10647 [hep-ph]} \BibitemShut
  {NoStop}%
\bibitem [{\citenamefont {Oudichhya}\ \emph
  {et~al.}(2021{\natexlab{b}})\citenamefont {Oudichhya}, \citenamefont
  {Gandhi},\ and\ \citenamefont {Rai}}]{Oudichhya:2021yln}%
  \BibitemOpen
  \bibfield  {author} {\bibinfo {author} {\bibfnamefont {J.}~\bibnamefont
  {Oudichhya}}, \bibinfo {author} {\bibfnamefont {K.}~\bibnamefont {Gandhi}}, \
  and\ \bibinfo {author} {\bibfnamefont {A.~K.}\ \bibnamefont {Rai}},\
  }\bibfield  {title} {\emph {\bibinfo {title} {{Mass-spectra of singly,
  doubly, and triply bottom baryons}},\ }}\href {\doibase
  10.1103/PhysRevD.104.114027} {\bibfield  {journal} {\bibinfo  {journal}
  {Phys. Rev. D}\ }\textbf {\bibinfo {volume} {104}},\ \bibinfo {pages}
  {114027} (\bibinfo {year} {2021}{\natexlab{b}})},\ \Eprint
  {http://arxiv.org/abs/2111.00236} {arXiv:2111.00236 [hep-ph]} \BibitemShut
  {NoStop}%
\bibitem [{\citenamefont {Jin}\ \emph {et~al.}(2013)\citenamefont {Jin},
  \citenamefont {Li}, \citenamefont {Si}, \citenamefont {Yang},\ and\
  \citenamefont {Yao}}]{Jin:2013bra}%
  \BibitemOpen
  \bibfield  {author} {\bibinfo {author} {\bibfnamefont {Y.}~\bibnamefont
  {Jin}}, \bibinfo {author} {\bibfnamefont {S.-Y.}\ \bibnamefont {Li}},
  \bibinfo {author} {\bibfnamefont {Z.-G.}\ \bibnamefont {Si}}, \bibinfo
  {author} {\bibfnamefont {Z.-J.}\ \bibnamefont {Yang}}, \ and\ \bibinfo
  {author} {\bibfnamefont {T.}~\bibnamefont {Yao}},\ }\bibfield  {title} {\emph
  {\bibinfo {title} {{Colour connections of four quark $Q\bar{Q}Q'\bar{Q}'$
  system and doubly heavy baryon production in $e^{+}e^{-}$ annihilation}},\
  }}\href {\doibase 10.1016/j.physletb.2013.10.070} {\bibfield  {journal}
  {\bibinfo  {journal} {Phys. Lett. B}\ }\textbf {\bibinfo {volume} {727}},\
  \bibinfo {pages} {468} (\bibinfo {year} {2013})},\ \Eprint
  {http://arxiv.org/abs/1309.5849} {arXiv:1309.5849 [hep-ph]} \BibitemShut
  {NoStop}%
\bibitem [{\citenamefont {Jiang}\ \emph {et~al.}(2013)\citenamefont {Jiang},
  \citenamefont {Wu}, \citenamefont {Wang}, \citenamefont {Zhang},\ and\
  \citenamefont {Fang}}]{Jiang:2013ej}%
  \BibitemOpen
  \bibfield  {author} {\bibinfo {author} {\bibfnamefont {J.}~\bibnamefont
  {Jiang}}, \bibinfo {author} {\bibfnamefont {X.-G.}\ \bibnamefont {Wu}},
  \bibinfo {author} {\bibfnamefont {S.-M.}\ \bibnamefont {Wang}}, \bibinfo
  {author} {\bibfnamefont {J.-W.}\ \bibnamefont {Zhang}}, \ and\ \bibinfo
  {author} {\bibfnamefont {Z.-Y.}\ \bibnamefont {Fang}},\ }\bibfield  {title}
  {\emph {\bibinfo {title} {{A Further Study on the Doubly Heavy Baryon
  Production around the $Z^0$ Peak at A High Luminosity $e^+ e^-$ Collider}},\
  }}\href {\doibase 10.1103/PhysRevD.87.054027} {\bibfield  {journal} {\bibinfo
   {journal} {Phys. Rev. D}\ }\textbf {\bibinfo {volume} {87}},\ \bibinfo
  {pages} {054027} (\bibinfo {year} {2013})},\ \Eprint
  {http://arxiv.org/abs/1302.0601} {arXiv:1302.0601 [hep-ph]} \BibitemShut
  {NoStop}%
\bibitem [{\citenamefont {Jiang}\ \emph {et~al.}(2012)\citenamefont {Jiang},
  \citenamefont {Wu}, \citenamefont {Liao}, \citenamefont {Zheng},\ and\
  \citenamefont {Fang}}]{Jiang:2012jt}%
  \BibitemOpen
  \bibfield  {author} {\bibinfo {author} {\bibfnamefont {J.}~\bibnamefont
  {Jiang}}, \bibinfo {author} {\bibfnamefont {X.-G.}\ \bibnamefont {Wu}},
  \bibinfo {author} {\bibfnamefont {Q.-L.}\ \bibnamefont {Liao}}, \bibinfo
  {author} {\bibfnamefont {X.-C.}\ \bibnamefont {Zheng}}, \ and\ \bibinfo
  {author} {\bibfnamefont {Z.-Y.}\ \bibnamefont {Fang}},\ }\bibfield  {title}
  {\emph {\bibinfo {title} {{Doubly Heavy Baryon Production at A High
  Luminosity $e^+ e^-$ Collider}},\ }}\href {\doibase
  10.1103/PhysRevD.86.054021} {\bibfield  {journal} {\bibinfo  {journal} {Phys.
  Rev. D}\ }\textbf {\bibinfo {volume} {86}},\ \bibinfo {pages} {054021}
  (\bibinfo {year} {2012})},\ \Eprint {http://arxiv.org/abs/1208.3051}
  {arXiv:1208.3051 [hep-ph]} \BibitemShut {NoStop}%
\bibitem [{\citenamefont {Zhang}\ \emph {et~al.}(2011)\citenamefont {Zhang},
  \citenamefont {Wu}, \citenamefont {Zhong}, \citenamefont {Yu},\ and\
  \citenamefont {Fang}}]{Zhang:2011hi}%
  \BibitemOpen
  \bibfield  {author} {\bibinfo {author} {\bibfnamefont {J.-W.}\ \bibnamefont
  {Zhang}}, \bibinfo {author} {\bibfnamefont {X.-G.}\ \bibnamefont {Wu}},
  \bibinfo {author} {\bibfnamefont {T.}~\bibnamefont {Zhong}}, \bibinfo
  {author} {\bibfnamefont {Y.}~\bibnamefont {Yu}}, \ and\ \bibinfo {author}
  {\bibfnamefont {Z.-Y.}\ \bibnamefont {Fang}},\ }\bibfield  {title} {\emph
  {\bibinfo {title} {{Hadronic Production of the Doubly Heavy Baryon $\Xi_{bc}$
  at LHC}},\ }}\href {\doibase 10.1103/PhysRevD.83.034026} {\bibfield
  {journal} {\bibinfo  {journal} {Phys. Rev. D}\ }\textbf {\bibinfo {volume}
  {83}},\ \bibinfo {pages} {034026} (\bibinfo {year} {2011})},\ \Eprint
  {http://arxiv.org/abs/1101.1130} {arXiv:1101.1130 [hep-ph]} \BibitemShut
  {NoStop}%
\bibitem [{\citenamefont {Ma}\ and\ \citenamefont {Si}(2003)}]{Ma:2003zk}%
  \BibitemOpen
  \bibfield  {author} {\bibinfo {author} {\bibfnamefont {J.~P.}\ \bibnamefont
  {Ma}}\ and\ \bibinfo {author} {\bibfnamefont {Z.~G.}\ \bibnamefont {Si}},\
  }\bibfield  {title} {\emph {\bibinfo {title} {{Factorization approach for
  inclusive production of doubly heavy baryon}},\ }}\href {\doibase
  10.1016/j.physletb.2003.06.064} {\bibfield  {journal} {\bibinfo  {journal}
  {Phys. Lett. B}\ }\textbf {\bibinfo {volume} {568}},\ \bibinfo {pages} {135}
  (\bibinfo {year} {2003})},\ \Eprint {http://arxiv.org/abs/hep-ph/0305079}
  {arXiv:hep-ph/0305079} \BibitemShut {NoStop}%
\bibitem [{\citenamefont {Chen}\ \emph {et~al.}(2014)\citenamefont {Chen},
  \citenamefont {Wu}, \citenamefont {Sun}, \citenamefont {Ma},\ and\
  \citenamefont {Fu}}]{Chen:2014frw}%
  \BibitemOpen
  \bibfield  {author} {\bibinfo {author} {\bibfnamefont {G.}~\bibnamefont
  {Chen}}, \bibinfo {author} {\bibfnamefont {X.-G.}\ \bibnamefont {Wu}},
  \bibinfo {author} {\bibfnamefont {Z.}~\bibnamefont {Sun}}, \bibinfo {author}
  {\bibfnamefont {Y.}~\bibnamefont {Ma}}, \ and\ \bibinfo {author}
  {\bibfnamefont {H.-B.}\ \bibnamefont {Fu}},\ }\bibfield  {title} {\emph
  {\bibinfo {title} {{Photoproduction of doubly heavy baryon at the ILC}},\
  }}\href {\doibase 10.1007/JHEP12(2014)018} {\bibfield  {journal} {\bibinfo
  {journal} {JHEP}\ }\textbf {\bibinfo {volume} {12}},\ \bibinfo {pages} {018}
  (\bibinfo {year} {2014})},\ \Eprint {http://arxiv.org/abs/1408.4615}
  {arXiv:1408.4615 [hep-ph]} \BibitemShut {NoStop}%
\bibitem [{\citenamefont {Yang}\ \emph {et~al.}(2014)\citenamefont {Yang},
  \citenamefont {Zhang},\ and\ \citenamefont {Zheng}}]{Yang:2014ita}%
  \BibitemOpen
  \bibfield  {author} {\bibinfo {author} {\bibfnamefont {Z.-J.}\ \bibnamefont
  {Yang}}, \bibinfo {author} {\bibfnamefont {P.-F.}\ \bibnamefont {Zhang}}, \
  and\ \bibinfo {author} {\bibfnamefont {Y.-J.}\ \bibnamefont {Zheng}},\
  }\bibfield  {title} {\emph {\bibinfo {title} {{Doubly Heavy Baryon Production
  in $e^{+}e^{-}$ Annihilation}},\ }}\href {\doibase
  10.1088/0256-307X/31/5/051301} {\bibfield  {journal} {\bibinfo  {journal}
  {Chin. Phys. Lett.}\ }\textbf {\bibinfo {volume} {31}},\ \bibinfo {pages}
  {051301} (\bibinfo {year} {2014})}\BibitemShut {NoStop}%
\bibitem [{\citenamefont {Saleev}(1999)}]{Saleev:1999ti}%
  \BibitemOpen
  \bibfield  {author} {\bibinfo {author} {\bibfnamefont {V.~A.}\ \bibnamefont
  {Saleev}},\ }\bibfield  {title} {\emph {\bibinfo {title} {{Omega(ccc)
  production via fragmentation at LHC}},\ }}\href {\doibase
  10.1142/S0217732399002741} {\bibfield  {journal} {\bibinfo  {journal} {Mod.
  Phys. Lett. A}\ }\textbf {\bibinfo {volume} {14}},\ \bibinfo {pages} {2615}
  (\bibinfo {year} {1999})},\ \Eprint {http://arxiv.org/abs/hep-ph/9906515}
  {arXiv:hep-ph/9906515} \BibitemShut {NoStop}%
\bibitem [{\citenamefont {Gomshi~Nobary}\ and\ \citenamefont
  {Sepahvand}(2006{\natexlab{a}})}]{GomshiNobary:2005ur}%
  \BibitemOpen
  \bibfield  {author} {\bibinfo {author} {\bibfnamefont {M.~A.}\ \bibnamefont
  {Gomshi~Nobary}}\ and\ \bibinfo {author} {\bibfnamefont {R.}~\bibnamefont
  {Sepahvand}},\ }\bibfield  {title} {\emph {\bibinfo {title} {{An Ivestigation
  of triply heavy baryon production at hadron colliders}},\ }}\href {\doibase
  10.1016/j.nuclphysb.2006.01.043} {\bibfield  {journal} {\bibinfo  {journal}
  {Nucl. Phys. B}\ }\textbf {\bibinfo {volume} {741}},\ \bibinfo {pages} {34}
  (\bibinfo {year} {2006}{\natexlab{a}})},\ \Eprint
  {http://arxiv.org/abs/hep-ph/0508115} {arXiv:hep-ph/0508115} \BibitemShut
  {NoStop}%
\bibitem [{\citenamefont {Gomshi~Nobary}\ and\ \citenamefont
  {Sepahvand}(2006{\natexlab{b}})}]{GomshiNobary:2006tzy}%
  \BibitemOpen
  \bibfield  {author} {\bibinfo {author} {\bibfnamefont {M.~A.}\ \bibnamefont
  {Gomshi~Nobary}}\ and\ \bibinfo {author} {\bibfnamefont {R.}~\bibnamefont
  {Sepahvand}},\ }\bibfield  {title} {\emph {\bibinfo {title} {{Triply heavy
  baryons}},\ }}\href@noop {} {\bibfield  {journal} {\bibinfo  {journal}
  {eConf}\ }\textbf {\bibinfo {volume} {C0605151}},\ \bibinfo {pages} {0010}
  (\bibinfo {year} {2006}{\natexlab{b}})},\ \Eprint
  {http://arxiv.org/abs/hep-ph/0609123} {arXiv:hep-ph/0609123} \BibitemShut
  {NoStop}%
\bibitem [{\citenamefont {Gomshi~Nobary}\ and\ \citenamefont
  {Sepahvand}(2005)}]{GomshiNobary:2004mq}%
  \BibitemOpen
  \bibfield  {author} {\bibinfo {author} {\bibfnamefont {M.~A.}\ \bibnamefont
  {Gomshi~Nobary}}\ and\ \bibinfo {author} {\bibfnamefont {R.}~\bibnamefont
  {Sepahvand}},\ }\bibfield  {title} {\emph {\bibinfo {title} {{Fragmentation
  of triply heavy baryons}},\ }}\href {\doibase 10.1103/PhysRevD.71.034024}
  {\bibfield  {journal} {\bibinfo  {journal} {Phys. Rev. D}\ }\textbf {\bibinfo
  {volume} {71}},\ \bibinfo {pages} {034024} (\bibinfo {year} {2005})},\
  \Eprint {http://arxiv.org/abs/hep-ph/0406148} {arXiv:hep-ph/0406148}
  \BibitemShut {NoStop}%
\bibitem [{\citenamefont {Chen}\ and\ \citenamefont {Wu}(2011)}]{Chen:2011mb}%
  \BibitemOpen
  \bibfield  {author} {\bibinfo {author} {\bibfnamefont {Y.-Q.}\ \bibnamefont
  {Chen}}\ and\ \bibinfo {author} {\bibfnamefont {S.-Z.}\ \bibnamefont {Wu}},\
  }\bibfield  {title} {\emph {\bibinfo {title} {{Production of Triply Heavy
  Baryons at LHC}},\ }}\href {\doibase 10.1007/JHEP08(2011)144} {\bibfield
  {journal} {\bibinfo  {journal} {JHEP}\ }\textbf {\bibinfo {volume} {08}},\
  \bibinfo {pages} {144} (\bibinfo {year} {2011})},\ \bibinfo {note} {[Erratum:
  JHEP 09, 089 (2011)]},\ \Eprint {http://arxiv.org/abs/1106.0193}
  {arXiv:1106.0193 [hep-ph]} \BibitemShut {NoStop}%
\bibitem [{\citenamefont {Wu}\ \emph {et~al.}(2012)\citenamefont {Wu},
  \citenamefont {Li},\ and\ \citenamefont {Rashidin}}]{Wu:2012wj}%
  \BibitemOpen
  \bibfield  {author} {\bibinfo {author} {\bibfnamefont {S.-Z.}\ \bibnamefont
  {Wu}}, \bibinfo {author} {\bibfnamefont {Y.-W.}\ \bibnamefont {Li}}, \ and\
  \bibinfo {author} {\bibfnamefont {R.}~\bibnamefont {Rashidin}},\ }\bibfield
  {title} {\emph {\bibinfo {title} {{Heaviest bound baryons production at the
  Large Hadron Collider}},\ }}\href {\doibase 10.1103/PhysRevD.86.114504}
  {\bibfield  {journal} {\bibinfo  {journal} {Phys. Rev. D}\ }\textbf {\bibinfo
  {volume} {86}},\ \bibinfo {pages} {114504} (\bibinfo {year}
  {2012})}\BibitemShut {NoStop}%
\bibitem [{\citenamefont {Zhao}\ and\ \citenamefont
  {Zhuang}(2017)}]{Zhao:2017gpq}%
  \BibitemOpen
  \bibfield  {author} {\bibinfo {author} {\bibfnamefont {J.}~\bibnamefont
  {Zhao}}\ and\ \bibinfo {author} {\bibfnamefont {P.}~\bibnamefont {Zhuang}},\
  }\bibfield  {title} {\emph {\bibinfo {title} {{Multicharmed Baryon Production
  in High Energy Nuclear Collisions}},\ }}\href {\doibase
  10.1007/s00601-017-1255-9} {\bibfield  {journal} {\bibinfo  {journal} {Few
  Body Syst.}\ }\textbf {\bibinfo {volume} {58}},\ \bibinfo {pages} {100}
  (\bibinfo {year} {2017})}\BibitemShut {NoStop}%
\bibitem [{\citenamefont {He}\ \emph {et~al.}(2015)\citenamefont {He},
  \citenamefont {Liu},\ and\ \citenamefont {Zhuang}}]{He:2014tga}%
  \BibitemOpen
  \bibfield  {author} {\bibinfo {author} {\bibfnamefont {H.}~\bibnamefont
  {He}}, \bibinfo {author} {\bibfnamefont {Y.}~\bibnamefont {Liu}}, \ and\
  \bibinfo {author} {\bibfnamefont {P.}~\bibnamefont {Zhuang}},\ }\bibfield
  {title} {\emph {\bibinfo {title} {{$\Omega_{ccc}$ production in high energy
  nuclear collisions}},\ }}\href {\doibase 10.1016/j.physletb.2015.04.049}
  {\bibfield  {journal} {\bibinfo  {journal} {Phys. Lett. B}\ }\textbf
  {\bibinfo {volume} {746}},\ \bibinfo {pages} {59} (\bibinfo {year} {2015})},\
  \Eprint {http://arxiv.org/abs/1409.1009} {arXiv:1409.1009 [hep-ph]}
  \BibitemShut {NoStop}%
\bibitem [{\citenamefont {Becattini}(2005)}]{Becattini:2005hb}%
  \BibitemOpen
  \bibfield  {author} {\bibinfo {author} {\bibfnamefont {F.}~\bibnamefont
  {Becattini}},\ }\bibfield  {title} {\emph {\bibinfo {title} {{Production of
  multiply heavy flavored baryons from quark gluon plasma in relativistic heavy
  ion collisions}},\ }}\href {\doibase 10.1103/PhysRevLett.95.022301}
  {\bibfield  {journal} {\bibinfo  {journal} {Phys. Rev. Lett.}\ }\textbf
  {\bibinfo {volume} {95}},\ \bibinfo {pages} {022301} (\bibinfo {year}
  {2005})},\ \Eprint {http://arxiv.org/abs/hep-ph/0503239}
  {arXiv:hep-ph/0503239} \BibitemShut {NoStop}%
\bibitem [{\citenamefont {Baranov}\ and\ \citenamefont
  {Slad}(2004)}]{Baranov:2004er}%
  \BibitemOpen
  \bibfield  {author} {\bibinfo {author} {\bibfnamefont {S.~P.}\ \bibnamefont
  {Baranov}}\ and\ \bibinfo {author} {\bibfnamefont {V.~L.}\ \bibnamefont
  {Slad}},\ }\bibfield  {title} {\emph {\bibinfo {title} {{Production of triply
  charmed Omega(ccc) baryons in e+ e- annihilation}},\ }}\href {\doibase
  10.1134/1.1707141} {\bibfield  {journal} {\bibinfo  {journal} {Phys. Atom.
  Nucl.}\ }\textbf {\bibinfo {volume} {67}},\ \bibinfo {pages} {808} (\bibinfo
  {year} {2004})},\ \Eprint {http://arxiv.org/abs/hep-ph/0603090}
  {arXiv:hep-ph/0603090} \BibitemShut {NoStop}%
\bibitem [{\citenamefont {Chang}\ and\ \citenamefont
  {Chen}(1992)}]{Chang:1992bb}%
  \BibitemOpen
  \bibfield  {author} {\bibinfo {author} {\bibfnamefont {C.-H.}\ \bibnamefont
  {Chang}}\ and\ \bibinfo {author} {\bibfnamefont {Y.-Q.}\ \bibnamefont
  {Chen}},\ }\bibfield  {title} {\emph {\bibinfo {title} {{The Production of
  B(c) or anti-B(c) meson associated with two heavy quark jets in Z0 boson
  decay}},\ }}\href {\doibase 10.1103/PhysRevD.46.3845} {\bibfield  {journal}
  {\bibinfo  {journal} {Phys. Rev. D}\ }\textbf {\bibinfo {volume} {46}},\
  \bibinfo {pages} {3845} (\bibinfo {year} {1992})},\ \bibinfo {note}
  {[Erratum: Phys.Rev.D 50, 6013(E) (1994)]}\BibitemShut {NoStop}%
\bibitem [{\citenamefont {Bodwin}\ \emph {et~al.}(1995)\citenamefont {Bodwin},
  \citenamefont {Braaten},\ and\ \citenamefont {Lepage}}]{Bodwin:1994jh}%
  \BibitemOpen
  \bibfield  {author} {\bibinfo {author} {\bibfnamefont {G.~T.}\ \bibnamefont
  {Bodwin}}, \bibinfo {author} {\bibfnamefont {E.}~\bibnamefont {Braaten}}, \
  and\ \bibinfo {author} {\bibfnamefont {G.~P.}\ \bibnamefont {Lepage}},\
  }\bibfield  {title} {\emph {\bibinfo {title} {{Rigorous QCD analysis of
  inclusive annihilation and production of heavy quarkonium}},\ }}\href
  {\doibase 10.1103/PhysRevD.55.5853} {\bibfield  {journal} {\bibinfo
  {journal} {Phys. Rev. D}\ }\textbf {\bibinfo {volume} {51}},\ \bibinfo
  {pages} {1125} (\bibinfo {year} {1995})},\ \bibinfo {note} {[Erratum:
  Phys.Rev.D 55, 5853(E) (1997)]},\ \Eprint {http://arxiv.org/abs/hep-ph/9407339}
  {arXiv:hep-ph/9407339} \BibitemShut {NoStop}%
\bibitem [{\citenamefont {Lepage}(1978)}]{Lepage:1977sw}%
  \BibitemOpen
  \bibfield  {author} {\bibinfo {author} {\bibfnamefont {G.~P.}\ \bibnamefont
  {Lepage}},\ }\bibfield  {title} {\emph {\bibinfo {title} {{A New Algorithm
  for Adaptive Multidimensional Integration}},\ }}\href {\doibase
  10.1016/0021-9991(78)90004-9} {\bibfield  {journal} {\bibinfo  {journal} {J.
  Comput. Phys.}\ }\textbf {\bibinfo {volume} {27}},\ \bibinfo {pages} {192}
  (\bibinfo {year} {1978})}\BibitemShut {NoStop}%
\bibitem [{\citenamefont {An}\ \emph {et~al.}(2019)\citenamefont {An} \emph
  {et~al.}}]{An:2018dwb}%
  \BibitemOpen
  \bibfield  {author} {\bibinfo {author} {\bibfnamefont {F.}~\bibnamefont {An}}
  \emph {et~al.},\ }\bibfield  {title} {\emph {\bibinfo {title} {{Precision
  Higgs physics at the CEPC}},\ }}\href {\doibase
  10.1088/1674-1137/43/4/043002} {\bibfield  {journal} {\bibinfo  {journal}
  {Chin. Phys. C}\ }\textbf {\bibinfo {volume} {43}},\ \bibinfo {pages}
  {043002} (\bibinfo {year} {2019})},\ \Eprint
  {http://arxiv.org/abs/1810.09037} {arXiv:1810.09037 [hep-ex]} \BibitemShut
  {NoStop}%
\bibitem [{\citenamefont {Zhang}\ \emph {et~al.}(2022)\citenamefont {Zhang},
  \citenamefont {Zheng},\ and\ \citenamefont {Wu}}]{Zhang:2021ypo}%
  \BibitemOpen
  \bibfield  {author} {\bibinfo {author} {\bibfnamefont {Z.-Y.}\ \bibnamefont
  {Zhang}}, \bibinfo {author} {\bibfnamefont {X.-C.}\ \bibnamefont {Zheng}}, \
  and\ \bibinfo {author} {\bibfnamefont {X.-G.}\ \bibnamefont {Wu}},\
  }\bibfield  {title} {\emph {\bibinfo {title} {{Production of the $B_c$ meson
  at the CEPC}},\ }}\href {\doibase 10.1140/epjc/s10052-022-10212-4} {\bibfield
   {journal} {\bibinfo  {journal} {Eur. Phys. J. C}\ }\textbf {\bibinfo
  {volume} {82}},\ \bibinfo {pages} {246} (\bibinfo {year} {2022})},\ \Eprint
  {http://arxiv.org/abs/2111.13917} {arXiv:2111.13917 [hep-ph]} \BibitemShut
  {NoStop}%
\end{thebibliography}
\end{document}